\documentclass[12pt]{article}
\usepackage{epsfig}
\usepackage{amsmath, amssymb, graphics, setspace}

\newcommand \bra[1]{\left< {#1} \,\right\vert}
\newcommand \ket[1]{\left\vert\, {#1} \, \right>}

\newcommand{\bea}{\begin{eqnarray}}
\newcommand{\eea}{\end{eqnarray}}
\newcommand{\simgt}{\hbox{ \raise3pt\hbox to 0pt{$>$}\raise-3pt\hbox{$\sim$} }}
\newcommand{\simlt}{\hbox{ \raise3pt\hbox to 0pt{$<$}\raise-3pt\hbox{$\sim$} }}

\newcommand{\clfn}{\setcounter{footnote}{0}}

\newcommand{\LQ}{\Lambda_{\rm QCD}}

\begin{document}

\begin{titlepage}

    \begin{flushright}
      \normalsize TU-980\\
      \today
    \end{flushright}

\vskip1.5cm
\begin{center}
\Large\bf\boldmath
Full Formula for Heavy Quarkonium Energy Levels\\
at Next-to-next-to-next-to-leading Order
\unboldmath
\end{center}

\vspace*{0.8cm}
\begin{center}

{\sc Y. Kiyo}$^{a}$ and
{\sc Y. Sumino$^{b}$}\\[5mm]
  {\small\it $^a$ Department of Physics, Juntendo University}\\[0.1cm]
  {\small\it Inzai, Chiba 270-1695, Japan}

  {\small\it $^b$ Department of Physics, Tohoku University}\\[0.1cm]
  {\small\it Sendai, 980-8578 Japan}

\end{center}

\vspace*{0.8cm}
\begin{abstract}
\noindent
We derive a full formula
for the energy level of a
heavy quarkonium state identified by the
quantum numbers $n$, $\ell$, $s$ and $j$,
up to ${\cal O}(\alpha_s^5 m)$
and ${\cal O}(\alpha_s^5 m \log \alpha_s)$ in perturbative QCD.
The QCD Bethe logarithm is given
in a one-parameter integral form.
The rest of the formula is given as a combination of
rational numbers, transcendental numbers 
($\pi$, $\zeta(3)$, $\zeta(5)$) and finite sums
(besides the 3-loop constant $\bar{a}_3$
of the static potential whose full analytic form is still unknown).
A derivation of the formula is given.

\vspace*{0.8cm}
\noindent
PACS numbers: 12.38.Bx, 12.15.Ff, 14.40.Pq

\end{abstract}

\vfil
\end{titlepage}

\section{Introduction}

For a long time
properties of various hadrons have been studied
in order to obtain a better understanding of
dynamics of the strong interaction.
Among various observed hadrons,
the heavy quarkonium states are unique, in the sense
that these are the
only known individual hadronic states whose
properties can be predicted in a self-contained manner
within perturbative QCD.
Namely, several observables associated with individual
heavy quarkonium states
(such as energy levels,  leptonic decay widths  and
transition rates) can be computed systematically in expansions in 
the strong coupling constant $\alpha_s$.
Such series expansions make sense,
since the large mass of
the heavy quarks,  $m(\gg \LQ$),
and the color-singlet nature of this bound state restricts the relevant
dynamical degrees of freedom of this system to 
those only in an ultraviolet (UV) region
and the asymptotic freedom of QCD designates
expansions in a small coupling constant.

To carry out computations of higher-order terms of these
expansions, 
development of effective field theories (EFTs) for heavy quarkonium
states, such as potential-NRQCD (pNRQCD)  \cite{Pineda:1997bj}
and velocity-NRQCD (vNRQCD) \cite{Luke:1999kz}, has been crucial.
It is also essential that
various computational technologies 
have made rapid progress during the same period.
(See e.g.\ \cite{Smirnov:2002pj,Smirnov:2004ym}.)
These theoretical tools enable systematic
organization of computations of higher-order corrections
by separating different energy
scales involved in the computations in a systematic manner \cite{Beneke:1997zp}.
These developments opened up vast applications of perturbative
QCD in heavy quarkonium physics. 
See \cite{Brambilla:2004wf} for comprehensive reviews. 
Some of more recent results 
can be found in  \cite{Penin:2004ay,Pineda:2006gx,
Kiyo:2010jm,Beneke:2014qea,Penin:2014zaa,Ayala:2014yxa,Bazavov:2014soa}.

Among the observables of heavy quarkonium states,
it turns out that the energy levels 
can be predicted particularly accurately in
perturbative QCD.
The decoupling of infrared (IR) degrees of freedom from these
states can be naturally incorporated in computations of
the energy levels by using 
a short-distance mass of the heavy quarks instead of the
pole mass, and this prescription improves convergence
of the perturbative series of the energy levels drastically
 \cite{Pineda:id}.

Spectroscopy of heavy quarkonium states (in particular
the bottomonium states) has provided an important testing ground
of perturbative QCD.
The full ${\cal O}(\alpha_s^4 m)$ and ${\cal O}(\alpha_s^4 m \log \alpha_s)$ 
corrections to the energy levels 
were computed in \cite{Pineda:1997hz}.
Analyses of the bottomonium spectrum, which incorporate the
IR decoupling and the perturbative corrections up to
this order, have shown that the gross structure
of the bottomonium spectrum, including the levels of the
$n=1,2$ and some of the $n=3$ states
($n$ is the principal quantum number), is reproduced reasonably
well within the estimated perturbative uncertainties
\cite{Brambilla:2001fw}.
Important applications of the spectroscopy are precise determinations
of the heavy quark masses from the lowest-lying energy levels.
The bottom and charm quark masses have been determined,
and in the future the top quark mass is expected to be determined
accurately in this way.

Stability of the predictions  for the energy levels and their agreement 
with experimental data  are predominantly determined by 
the prediction for the static energy
$E_{\rm stat}(r)=2m_{\rm pole}+V_{\rm QCD}(r)$.
 Once the pole mass is expressed in terms of a short-distance mass,
the prediction for $E_{\rm stat}(r)$ in perturbative QCD
agrees with lattice computations or typical phenomenological potentials
in the relevant distance range
\cite{Sumino:2001eh,Pineda:2002se}.
Furthermore, by increasing the order of the perturbative expansion, 
the range of $r$, where
convergence and agreement are seen, becomes wider \cite{Anzai:2009tm}.

After the development of EFTs, computations of the ${\cal O}(\alpha_s^5 m )$
and ${\cal O}(\alpha_s^5 m \log\alpha_s)$
 corrections to the energy levels made progress.
Within pNRQCD, the corrections consist of two parts,
the contributions from the potential region and ultra-soft region.
The next-to-next-to-next-to-leading order
(NNNLO) Hamiltonian, which dictates the
contributions from the potential region, was computed in \cite{Kniehl:2002br}
(besides the 3-loop corrections 
to $V_{\rm QCD}(r)$, $\bar{a}_3$, which were computed later numerically in
\cite{Smirnov:2008pn,Anzai:2009tm,Smirnov:2009fh}).
The contributions from the ultra-soft region contain, besides the
part calculable analytically, 
a QCD analogue of the Bethe logarithm for the Lamb shift in QED.
The QCD Bethe logarithm for each state can be written as a 
one-parameter integral of 
elementary functions \cite{Kniehl:1999ud}.
Up to now, the ${\cal O}(\alpha_s^5 m \log \alpha_s)$ correction for a general  state
labeled by the quantum numbers
$(n,l,s,j)$ was computed in \cite{Brambilla:1999xj}, while 
the ${\cal O}(\alpha_s^5 m)$ and ${\cal O}(\alpha_s^5 m \log \alpha_s)$
corrections for a general $S$-wave state  
$(n,j)$  were computed in \cite{Beneke:2005hg}.
(See also \cite{Kiyo:2000fr} for earlier computations
of the ${\cal O}(\alpha_s^5 m)$ 
corrections.)

In this paper, 
we derive a full formula
for the energy level of a
heavy quarkonium state identified by the
quantum numbers $n$, $\ell$, $s$ and $j$,
up to ${\cal O}(\alpha_s^5 m)$
and ${\cal O}(\alpha_s^5 m \log \alpha_s)$.
The QCD Bethe logarithm is given
in a one-parameter integral form.
The rest of the formula is given as a combination of
rational numbers, transcendental numbers 
($\pi$, $\zeta(3)$, $\zeta(5)$) and finite sums
(besides $\bar{a}_3$ whose full analytic form is still unknown).
We explain details of the derivation of the formula.

We have already applied the formula to an analysis
of the bottomonium spectrum \cite{Kiyo:2013aea}.
The current status of the perturbative prediction
for the bottomonium spectrum
is practically determined by a fine-level cancellation
between $2m_{\rm pole}$ and $V_{\rm QCD}(r)$ in $E_{\rm stat}(r)$
and depends sensitively on the (yet unknown)
precise value of 
the order $\alpha_s^4$ correction to the
pole-$\overline{\rm MS}$ mass relation ($d_3$).
If $d_3$ is tuned to improve convergence of $E_{\rm stat}(r)$
maximally,
we observe a reasonable agreement between the predictions 
and experimental data of the bottomonium
spectrum within estimated perturbative
uncertainties.
On the other hand, the prediction becomes unstable
quickly if $d_3$ deviates from the fine-tuned value.
Hence, the status of the perturbative QCD prediction
at NNNLO is not yet clear until $d_3$ is computed.

This paper is organized as follows.
In Sec.~2 we explain a method to convert a class of
infinite
sums to combinations of transcendental numbers and
finite sums.
In Sec.~3 we explain briefly the theoretical framework
(pNRQCD) used in our computation of the energy levels.
The computation of NNNLO corrections from the potential
region is given in Sec.~4.
The computation of NNNLO corrections involving the ultra-soft region
is presented in Sec.~5.
The obtained general formula is presented in Sec.~6.
Sec.~7 is devoted to the summary and discussion.
Technical details are collected in the Appendices.
We set up our conventions and notations in App.~A.
Generating functions used for Fourier transformations
are given in App.~B.
Formulas useful for evaluating matrix elements
in terms of Coulomb wave functions are given in App.~C.
Evaluation of an expectation value of a spin-tensor operator
is explained in App.~D.
An arithmetic method for deriving a local Hamiltonian in the
ultra-soft correction is given in App.~E.
Methods for taking an angular average of matrix elements
for the ultra-soft corrections are given in App.~F.

\section{Finite Sum Formula at NNLO}
\label{sec:NNLO}

The known formula \cite{Pineda:1997hz} 
for the NNLO correction to the energy level 
includes an
infinite sum:
\bea
&&
A(n,\ell)=
\sum_{k=1}^{\infty}
\frac{(n-\ell+k-1)!}{(n+\ell+k)! \, k^3} .
\label{A}
\eea
For a given $n,\ell$ it is easily converted to a
combination of transcendental numbers and rational numbers.
As far as we know, however, a corresponding 
expression for arbitrary $n,\ell$ has not
been known.
We explain a method to
convert the above infinite sum to a combination of transcendental numbers 
[$\zeta(2)=\pi^2/6$ and $\zeta(3)$], rational numbers and
a finite sum.

By partial fractioning, one may write
\bea
&&
\frac{(n-\ell+k-1)!}{(n+\ell+k)!}=\prod_{m=-\ell}^{\ell}\frac{1}{n+k+m}
=\sum_{m=-\ell}^{\ell}\frac{R(\ell,m)}{n+k+m},
\label{partialfrac}
\eea
where
\bea
R(\ell,m)=\frac{(-1)^{\ell-m}}{(\ell+m)!(\ell-m)!}
\, .
\eea
Using the summation program \texttt{Wa} \cite{fn21}, which utilizes
the summation algorithm developed in \cite{Anzai:2012xw}, one may obtain a relation
\bea
\sum_{k=1}^\infty \frac{1}{(k+i)\,k^3}=\frac{\zeta(3)}{i}-\frac{\zeta(2)}{i^2}+\frac{S_1(i)}{i^3}
\, ,
\label{UseOfWa}
\eea
where
$S_1(i)=\sum_{k=1}^i\frac{1}{k}$
denotes the harmonic sum.
Furthermore, for a non-negative integer $a$, one may use the relation
\bea
&&
\sum_{m=-\ell}^{\ell}\frac{R(\ell,m)}{(n+m)^{a+1}}
=\frac{1}{a!}
\left( - \frac{\partial}{\partial k}\right)^a 
\sum_{m=-\ell}^{\ell}\frac{R(\ell,m)}{n+k+m}
\Biggr|_{k=0}
\nonumber
\\&&
~~~~~~~~~~~~~~~~~~~~~
=\frac{1}{a!}
\left( - \frac{\partial}{\partial k}\right)^a 
\frac{\Gamma(n-\ell+k)}{\Gamma(n+\ell+k+1)}
\Biggr|_{k=0}
\, .
\label{simplification}
\eea
Combining Eqs.~(\ref{A}), (\ref{partialfrac}), (\ref{UseOfWa}) and 
(\ref{simplification}),
one obtains an expression
\bea
&&
A(n,\ell)=
\frac{(n+\ell)!}{(n-\ell-1)!}
\left[\zeta(3)-\zeta(2)\,\Bigl\{S_1(n+\ell)-S_1(n-\ell-1)\Bigr\}
\right]
\nonumber
\\&&
~~~~~~~~~~~
+ \sum_{m=-\ell}^{\ell}\frac{R(\ell,m)}
{(n+m)^3}\,S_1(n+m)
\, ,
\eea
which includes only transcendental numbers and
finite sums.

This type of technique to convert an infinite sum to a combination of
finite sums and transcendental numbers is useful in simplifying 
the formula for the spectrum at NNNLO.

\section{Theoretical Framework}

In this section we explain the theoretical framework
which we use to compute the heavy quarkonium energy levels.

\subsection{Potential-NRQCD effective field theory}

pNRQCD \cite{Pineda:1997bj} is an EFT, which describes the interactions among
$Q\bar{Q}$ bound states and IR gluons and light quarks.
Color-singlet and octet $Q\bar{Q}$ composite states are
represented by the fields which are
local in time and bilocal in space coordinates:
\bea
S(\vec{x},\vec{r};t)_{\sigma\bar{\sigma}},~~~~~~~~~
{O}^a(\vec{x},\vec{r};t)_{\sigma\bar{\sigma}},
\eea
where
the spatial coordinates (2-component
spin indices) of $Q$ and $\bar{Q}$ are denoted by
$\vec{x}\pm \frac{1}{2}\vec{r}$ ($\sigma$ and $\bar{\sigma}$).
In addition, IR gluons and
quarks are included as dynamical degrees of freedom.
More precisely, pNRQCD 
is obtained after integrating out the hard (h)
and soft (s) modes of the dynamical fields in full QCD
\cite{Beneke:1997zp}:
\bea
&&
{\rm (h)}:~~~ p^0,|\vec{p}| \sim m ,
\\ &&
{\rm (s)}:~~~ p^0,|\vec{p}| \sim \alpha_s m .
\eea
The dynamical fields in pNRQCD have energy-momentum
in the potential (p) and 
ultra-soft (us) regions:
\bea
&&
{\rm (p)}:~~~ p^0\sim \alpha_s^2m,~~~|\vec{p}| \sim \alpha_s m ,
\label{PotentialRegion}
\\ &&
{\rm (us)}:~~~ p^0,|\vec{p}| \sim \alpha_s^2 m .
\eea
The details of the EFT derivation by the mode integration 
can be found in Ref.\cite{Beneke:2013jia}.

The Lagrangian of pNRQCD is given as expansions in $\vec{r}$ and
$1/m$:
\bea
{\cal L}_{\rm pNRQCD}=S^\dagger 
\bigl( i\partial_t - H_S \bigr) S +
{O}^{a\dagger}
\bigl(iD_t-H_{O} \bigr)^{ab}
{O}^b
+ g \, S^\dagger \, \vec{r}\cdot\vec{E}^a\, {O}^a
+\dots .
\label{pNRQCD-Lagrangian}
\eea
$H_S$ and $H_{O}$ denote
the quantum mechanical Hamiltonians
for the singlet and octet states, respectively, which 
are expressed by $\vec{x},\vec{r},\vec{\partial}_{x},\vec{\partial}_{r}$
and spin operators of $Q$ and $\bar{Q}$.
They represent the contributions from the potential region,
given as instantaneous interactions (i.e., potentials), 
 between $Q$ and $\bar{Q}$, and
dictate the main binding dynamics of the $Q\bar{Q}$ states.

The ultra-soft gluon fields are expanded in $\vec{r}$ (multipole expansion)
and expressed in terms of $A_\mu^a(\vec{x},t)$ and its derivatives.
The leading interaction of the singlet field $S$ and the 
ultra-soft gluon is given by a  dipole interaction
[the last terms of Eq.~(\ref{pNRQCD-Lagrangian})],
where the color-electric field is given by
$\vec{E}^a=
-\vec{\partial}_x A_0^a
-\partial_t \vec{A}^a-gf^{abc}{A}_0^b\vec{A}^c$.

\begin{figure}
\begin{center}
\includegraphics[width=9cm]{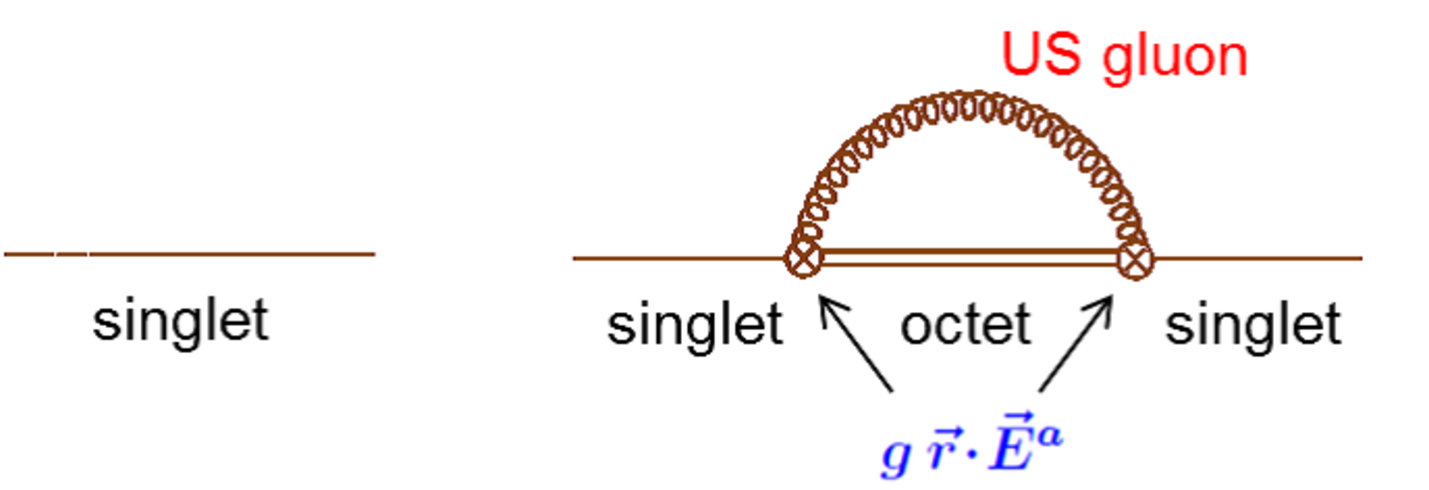}
\end{center}
\vspace*{-.5cm}
\caption{\small
First two diagrams for the full propagator of $S$
in multipole expansion in pNRQCD.
\label{Diagrams-pNRQCD}}
\end{figure}

The energy levels of the heavy quarkonium states are given as
the positions of poles of the full propagator of the
singlet field $S$.
The full propagator, in multipole expansion in $\vec{r}$,
is given by the diagrams in Fig.~\ref{Diagrams-pNRQCD}.
In momentum space and in the center-of-mass (c.m.)\ frame,\footnote{
It is obtained by setting the energy-momentum of the
center-of-gravity as ${P}^\mu=(2m+E,\vec{0})$ and
the initial and final relative momenta of 
$Q$ and $\bar{Q}$ as $\vec{p}$ and $\vec{p}^{\,\prime}$,
respectively. 
}
it is given by
\bea
&&
{\rm F.T.}\,
\bra{0}T\,S(\vec{x},\vec{r};t)S^\dagger (\vec{x}',\vec{r}^{\,\prime};t')\ket{0}
\biggr|_{\rm c.m.\,\, frame}
\nonumber\\&&
=\bra{\vec{p}}\frac{1}{E-H_S+i0}\ket{\vec{p}^{\,\prime}}
\nonumber\\&&
~~~
-ig^2 
\frac{T_F}{N_C}\int_0^\infty \! \! \! dt \, e^{iEt}\,
\bra{\vec{p}}\frac{1}{E-H_S+i0}
\, \vec{r}\!\cdot\!\vec{E}^a\!(t)
\,U^{ab}_O(t)
\,\vec{r}\!\cdot\!\vec{E}^b\!(0)
\frac{1}{E-H_S+i0}
\ket{\vec{p}^{\,\prime}}
\nonumber\\&&
~~~
+\cdots ,
\label{FullPropagatorS}
\eea
where $U_O(t)^{ab}$ denotes the octet propagator defined by
\bea
\bigl(iD_t-H_{O} \bigr)^{ab}
U^{bc}_O(t)=i\delta^{ac}\,\delta(t).
\label{DefOctetProp}
\eea
The first term of Eq.~(\ref{FullPropagatorS})
includes only contributions from the potential
region, whereas the second term includes contributions
from both potential 
and ultra-soft regions.

Since $|\vec{r}|$ is of the order of the Bohr radius,
$r\sim 1/(\alpha_s m)$, and the energy-momentum of
an ultra-soft gluon is $k^0,k\sim \alpha_s^2m$, the
multipole expansion (such as $\vec{r}\cdot \vec{\partial}A_0^a$,
$\vec{r}\cdot\partial_t\vec{A}^a$) generates expansion in
$\alpha_s$.
The second diagram of
Fig.~\ref{Diagrams-pNRQCD} is counted as NNNLO as there are two
dipole interactions and the associated couplings $g^2\sim\alpha_s$.

We decompose the 
Hamiltonian for the singlet state (in its c.m.\ frame)
as 
\begin{eqnarray}
H_S &=&
\frac{\hat{\vec{p}}^{\, 2}}{m}+ V_{\rm C}(r) + 
V_{\rm NC}(\hat{\vec{p}}, \vec{r}) 
~~~~~;~~~~~
\hat{\vec{p}}=-i\vec{\partial}_{r} \, ,
\label{SingletHamiltonian}
\eea
where $V_{\rm C}(r)$ and $V_{\rm NC}(\hat{\vec{p}}, \vec{r})$, respectively,
denote the Coulomb and non-Coulomb potentials.
Here and hereafter,
$m$ represents the pole mass of $Q$ or $\bar{Q}$.
Below we present each part of the potentials, which contributes to the
NNNLO corrections to the quarkonium energy levels.
These are determined in perturbative QCD by integrating
out the hard and soft modes.

For convenience, we use both coordinate-space
and momentum-space representations of the potentials.
In coordinate space we write each term of the potentials in the form
\bea
f(\hat{\vec{p}})\, V(\vec{r})\, g(\hat{\vec{p}}) ,
\eea
where $\hat{\vec{p}}=-i\vec{\partial}_{r}$ is
a derivative operator.
The corresponding potential in momentum space is
represented by
\bea
f(\vec{p})\, \widetilde{V}(\vec{q})\, g(\vec{p}^{\,\prime}),
\eea
with $\vec{p}^{\,\prime}=\vec{p}+\vec{q}$,
and
\bea
V(\vec{r}) =
\bar{\mu}^{2\epsilon}
\int\! \frac{d^d \vec{q}}{(2\pi)^d}\, e^{i \vec{q}\cdot \vec{r}} \,
\widetilde{V}(\vec{q}) 
~~~~~;~~~~~
\bar{\mu}^2\equiv\frac{\mu^2 e^{\gamma_E} }{4\pi} 
.
\eea
We employ dimensional regularization with one temporal
dimension and $d=D-1=3-2\epsilon$ spatial dimensions.
$\gamma_E=0.5772\cdots$ denotes the Euler constant.
The relation between both representations is given
explicitly by
\bea
&&
\bra{\vec{r}}f(\vec{p})V(\vec{r})g(\vec{p})\ket{\vec{r}^{\,\prime}}
=f(\hat{\vec{p}})\, V(\vec{r})\, g(\hat{\vec{p}})
\delta(\vec{r}-\vec{r}^{\,\prime})
,
\\&&
\bra{\vec{p}}f(\vec{p})V(\vec{r})g(\vec{p})\ket{\vec{p}^{\,\prime}}
=\bar{\mu}^{2\epsilon}
f(\vec{p})\, \widetilde{V}(\vec{q})\, g(\vec{p}^{\,\prime})
,
\\&&
\int d^d\vec{r} d^d\vec{r}^{\,\prime}\,
e^{i\vec{p}^{\,\prime}\!\cdot\vec{r}^{\,\prime}-i\vec{p}\cdot\vec{r}}
\bra{\vec{r}}f(\vec{p})V(\vec{r})g(\vec{p})\ket{\vec{r}^{\,\prime}}
=
\bra{\vec{p}}f(\vec{p})V(\vec{r})g(\vec{p})\ket{\vec{p}^{\,\prime}}
.
\eea

\subsection{Coulomb potential}
In purely perturbative expansion in $\alpha_s$,
the Coulomb potential $V_{\rm C}(r)$ is identical to
the static QCD potential, which is defined from the
expectation value of the Wilson loop and subtracting the
ultra-soft contribution.
At order $\alpha_s^4$ and beyond,
$V_{\rm C}(r)$ is IR divergent.
Generally we separate $V_{\rm C}(r)$ to
the renormalized part $V_{\rm C,ren}(r)$ and IR divergent part
$V_{\rm C,div}(r)$.
In the
computation of ultra-soft corrections,
$V_{\rm C,div}(r)$ plays the role of a counter term \cite{Pineda:1997bj} .

We define the separation of $1/\epsilon$-pole and the renormalized part in 
momentum space to conform with the conventional $\overline{\rm MS}$ scheme.
This is also so for the ultra-soft part for consistency [see Eqs. (\ref{HUSdiv}) and (\ref{HUSren})]. 
In momentum space, the Coulomb potential
up to order $\alpha_s^4$ is given by
\begin{eqnarray}
&&
\widetilde{V}_{\rm C}=\widetilde{V}_{\rm C,ren}+
\widetilde{V}_{\rm C,div} \, ,
\\
&&
\widetilde{V}_{\rm C,ren}(\vec{q})
=-4\pi C_F \frac{\alpha_{V,{\rm ren}}(q) }{q^2} ,
~~~~~~
\widetilde{V}_{\rm C,div}(\vec{q})=
-\frac{C_FC_A^3\alpha_s^4}{6 q^2}\,\frac{1}{{\epsilon}} ,
~~~~~;~~~~~
q=|\vec{q}| .
\end{eqnarray}

The $V$-scheme coupling constant is given as a perturbative 
expansion in the $\overline{\rm MS}$ coupling constant
$\alpha_s\equiv \alpha_s(\mu)$ as 
\begin{eqnarray}
&&
\alpha_{V,{\rm ren}} (q) = \alpha_s \sum_{n=0}^\infty P_n^V(L_q) \left(\frac{\alpha_s}{4\pi} \right)^n,
~~~~~~
L_q = \log \left(\frac{\mu^2}{q^2} \right),
\end{eqnarray}
where the coefficients of logarithms in $P_n^V$ are 
determined by the renormalization group as
\begin{eqnarray}
P^V_0&=& a_0,
\\
P^V_1 &=& a_1+ a_0\beta_0 L_q,
\nonumber
\\
P^V_2 &=& a_2 +\left(2a_1\beta_0+a_0\beta_1 \right) L_q+a_0\beta_0^{\, 2} L_q^{\, 2},
\nonumber
\\
P^V_3 &=& \bar{a}_3 +8\pi^2 C_A^3 L_q
+\left(3a_2\beta_0+2a_1 \beta_1+a_0\beta_2 \right) L_q 
\nonumber\\
&&
 +\left(3a_1\beta_0^{\, 2}+\frac{5}{2}a_0 \beta_0\beta_1\right) L_q^{\, 2} +a_0\beta_0^{\, 3} L_q^{\, 2}.
\end{eqnarray}
The $C_A^{\, 3}$ term in $P_3^{V}$ is a logarithm generated from  a combination of 
the $1/\epsilon$ pole of $\widetilde{V}_{\rm C, div}$ and  $({\mu}^{2\epsilon})^{3}$ 
of the three-loop integration measure in dimensional regularization.
The parameters $a_0,a_1,a_2,\bar{a}_3$ and $\beta_0,\beta_1,\beta_2$
are given in App.~\ref{AppCFandParams}.
The renormalized Coulomb potential in coordinate space is given by
\begin{eqnarray}
V_{\rm C,ren}(r)
&=& 
-\frac{C_F\alpha_s}{r} 
\bigg\{
a_0
+\left(\frac{\alpha_s}{4\pi}\right)
  \bigg[a_1 + a_0\beta_0 L_r \bigg]
\nonumber\\
&&
+\left(\frac{\alpha_s}{4\pi}\right)^2
  \bigg[a_2 
  +(2a_1\beta_0+a_0\beta_1) L_r 
  + a_0\beta_0^2 \bigg( L_r^{\, 2} +\frac{\pi^2}{3} \bigg) \bigg]
\nonumber\\
&&
+\left(\frac{\alpha_s}{4\pi}\right)^3
  \bigg[
\bar{a}_3 + 8\pi^2 C_A^{\,3} L_r
+ \left(3a_2\beta_0 +2a_1\beta_1+a_0\beta_2\right)L_r
\nonumber\\
&&\hspace{2cm}
+ \left(3a_1\beta_0^{\, 2} +\frac{5}{2}a_0\beta_0 \beta_1\right)  
   \bigg( L_r^{\, 2}+\frac{\pi^2}{3}\bigg)
\nonumber\\
&&\hspace{2cm}
  + a_0\beta_0^{\, 3} \bigg( L_r^{\, 3}+\pi^2L_r+16\zeta_3\bigg)
 \, \bigg]
\, \bigg\},
\label{V_{c,ren}(r)}
\end{eqnarray}
where $L_r = \log (e^{2\gamma_E} \mu^2 r^2)$.
For the renormalized part, Fourier transformation formula in $d=3$ can be
applied.

\subsection{Non-Coulomb potentials}
The non-Coulomb potentials also contain IR divergences,
and they are separated to the renormalized and divergent parts
in the same manner as the Coulomb potential.
The NNNLO  non-Coulomb potentials were obtained in Ref.~\cite{Kniehl:2002br}. 
In momentum space the renormalized non-Coulomb potentials are
given as follows.

\begin{eqnarray}
\widetilde{V}_{\rm NC,ren} &=&
(2\pi)^3 \delta(\vec{q}) \bigg( -\frac{\vec{p}^{\, 4}}{4m^3}\bigg)
+\frac{(4\pi C_F \alpha_s )\pi^2}{m q} C_{1/m}
\nonumber\\
&&
+\frac{\pi C_F \alpha_s}{m^2} 
\bigg( C_\delta + C_p \frac{\vec{p}^{\, 2}+\vec{p}^{\, \prime 2}}{2q^2} 
 +C_s \vec{S}^2 + C_\lambda \Lambda + C_t T \bigg),
\end{eqnarray}
where 
\begin{eqnarray}
\vec{S} &=& \frac{ \vec{\sigma}_1 +\vec{\sigma}_2}{2},
\\
\Lambda (\vec{p}, \vec{q})
&=&
i \frac{ \vec{S} \cdot (\vec{p}\times \vec{q}) }{ q^2},
\\
T(\vec{q}) &=& \vec{\sigma}_1\cdot  \vec{\sigma}_2
-3\frac{(\vec{q}\cdot \vec{\sigma}_1 )( \vec{q}\cdot \vec{\sigma}_2)}{q^2}.
\end{eqnarray}

We parameterize the potential coefficients as follows:
\begin{eqnarray}
C_X &=&\sum_{n=0}^\infty 
\left(\frac{\alpha_s}{4\pi}\right)^n d^X_{n}(q)
=
\sum_{n,k=0}^\infty 
\left(\frac{\alpha_s}{4\pi}\right)^n 
\left(L_q\right)^k
d^X_{nk} 
\end{eqnarray}
The $1/m$ potential coefficients at tree, one-loop and
two-loop levels are
given by
\begin{eqnarray}
d^{1/m}_0(q) &=& 0,
\\
d^{1/m}_1 (q)
&=& \frac{C_F}{2}-C_A ,
\\
d^{1/m}_2(q) 
&=&
4\bigg[
-\left(\frac{101}{36}+\frac{4}{3}\log{2}\right)C_A^2
+\left(\frac{65}{18}-\frac{8}{3}\log{2} \right)C_A C_F
+\frac{49}{36} C_A T_F n_l
\nonumber\\
&&
-\frac{2}{9} C_F T_F n_l
-\frac{4}{3}\left(C_A^2+2C_A C_F\right)L_q
\bigg]
+2 \beta_0 d_1^{1/m} L_q.
\label{eq:dcoef-1/m}
\end{eqnarray}
The $1/m^2$ potential coefficients at tree level
are given by
\begin{eqnarray}
&&
d^{\delta}_0 = 0,
\hspace{1cm} 
d^{p}_0 = -4,
\hspace{1cm} 
d^{s}_0 =  \frac{4}{3},
\hspace{1cm} 
d^{\lambda}_0 = 6,
\hspace{1cm} 
d^{t}_0 = \frac{1}{3},
\end{eqnarray}
and at one-loop level by
\begin{eqnarray}
d^{\delta}_1 (q)
&=& 
4 \bigg[
\left( 2C_A +2C_F -\frac{4}{15}T_F\right)
\nonumber\\
&&~
+\left(-\frac{17}{6} C_A -\frac{1}{3}C_F \right) L_m
+\bigg(\frac{25}{6}C_A-\frac{7}{3} C_F \bigg)L_q
\bigg]
+\beta_0 \, d^{\delta}_0 L_q
, \label{eq:dcoef-delta}
\\
d^{p}_1(q) 
&=&
4 \bigg[
\left( -\frac{31}{9}C_A +\frac{20}{9} T_F n_l \right)
 -\frac{8}{3} C_A L_q
\bigg]
+\beta_0 \, d^p_0  L_q
, \label{eq:dcoef-p}
\\ 
d^{s}_1 (q)
&=&
4 \bigg[
\left( \frac{22}{27}C_A -\frac{2}{3} C_F -\frac{20}{27} T_F n_l \right)
 + \frac{7}{6} C_A (L_m -L_q)
 \bigg]
+\beta_0\, d^s_0 L_q
, \label{eq:dcoef-s}   
\\
d^{\lambda}_1(q)  &=&
4\bigg[
 \left( \frac{31}{6} C_A+4 C_F -\frac{10}{3} T_F n_l \right)
+2C_A (L_m-L_q)
\bigg]
+\beta_0\, d^\lambda_0 L_q
, \label{eq:dcoef-lambda}   
\\
d^t_1(q)&=&
4 \bigg[
\left(  \frac{49}{108} C_A+\frac{1}{3} C_F -\frac{5}{27} T_F n_l \right)
+\frac{1}{6}C_A (L_m-L_q)
\bigg]
+\beta_0\, d^{t}_0 L_q
,
\label{eq:dcoef-t}   
\end{eqnarray}
where $L_m=\log(\mu^2/m^2)$ and $L_q=\log(\mu^2/q^2)$.
There are contributions from annihilation diagrams,
\begin{eqnarray}
&&
d^{\delta,a}_0 =0,
\hspace{1cm}
d^{\delta,a}_1 = 4 \left(-4+4\log{2}-2i\pi\right)T_F + \beta_0 d^{\delta,a}_0 L_m
, \label{eq:dcoef-delta-a}
\\
&&
d^{s,a}_0 =0,
\hspace{1cm}
d^{s,a}_1 = 4 \left(2-\log{2}+ i\pi\right)T_F + \beta_0 d^{s, a}_0 L_m
,
\label{eq:dcoef-s-a}
\end{eqnarray}
which should be added to $d_i^\delta$ and $d_i^s$.
The absorptive part is discarded in the computation of the energy levels.

In coordinate space the renormalized non-Coulomb potentials
are given as follows.
\bea
&&
V_{\rm NC,ren}(\hat{\vec{p}},\vec{r})=
V_{p^4}+V_{1/r^2}+V_\delta+V_{\{p^2\!,\,1/r\}}+
V_S+V_\Lambda+V_T 
+V_{\delta,a}+V_{S,a}
,
\label{RenormalizedNonCoulPot}
\\ &&
V_{p^4}= -\frac{\hat{\vec{p}}^{\, 4}}{4m^3},
\\ &&
V_{1/r^2}=
\frac{\alpha_s^2}{m r^2} 
\biggl[\,
\frac{1}{4}C_F(C_F-2 C_A)+\frac{\alpha_s}{\pi}C_F
\bigg\{ -\frac{2}{3}C_A(C_A+2C_F)(L_r+\log 2)
\nonumber\\&&
~~~~~~~~~~
+\frac{130C_AC_F-101C_A^2+(49C_A-8C_F)T_F n_l}{72}
+\frac{\beta_0(C_F-2C_A)}{8}L_r
\bigg\}\bigg] ,
\\ &&
V_\delta=\frac{\alpha_s^2}{m^2}C_F\biggl[\,
\frac{25C_A-14C_F}{12\pi}\,{\rm reg}\biggl[ \frac{1}{r^3}\biggr]
\nonumber\\&&
~~~~~~~~~~~~~~~~~~
-\frac{1}{30}
\{ 10C_F(L_m-6)+C_A(85L_m-60)+8T_F\}\delta(\vec{r})
\biggr],
\\&&
V_{\{p^2\!,\,1/r\}}=-\frac{C_F\alpha_s}{72m^2}
\biggl\{ \hat{\vec{p}}^{\,2},
\frac{1}{r}\Bigl[36+\frac{\alpha_s}{\pi}(C_A(24L_r+31)-20T_Fn_l+9\beta_0L_r)
\Bigr]\biggr\},
\\ &&
V_S=\frac{\vec{S}^2}{m^2}
\biggl[\, \pi C_F \alpha_s\delta(\vec{r})
\Bigl\{\frac{4}{3}+\frac{\alpha_s}{54\pi}(44C_A-36C_F+63C_AL_m-40T_Fn_l)
\Bigr\}
\nonumber\\&&
~~~~~~~~~~
+\frac{C_F\alpha_s^2}{\pi}\,{\rm reg}\biggl[ \frac{1}{r^3}\biggr]\,
\Bigl( -\frac{7}{12}C_A + \frac{1}{6}\beta_0
\Bigr) \biggr] ,
\\ &&
V_\Lambda= \delta_{\ell\ge 1}\,\frac{\vec{L}\cdot\vec{S}}{m^2r^3}\,
C_F\alpha_s\biggl[\,\frac{3}{2}+\frac{\alpha_s}{\pi}
\Bigl\{\frac{1}{24}(C_A(55+12L_m-12L_r)+4(6C_F-5T_Fn_l))
\nonumber\\&&
~~~~~~~~~~
+\frac{3}{8}\beta_0(L_r-2)\Bigr\}\biggr] ,
\\ &&
V_T=\delta_{\ell\ge 1}\,\frac{1}{m^2r^3}\biggl(
3\frac{(\vec{r}\cdot\vec{S})^2}{r^2}-\vec{S}^2\biggr)\,
C_F\alpha_s\biggl[\,\frac{1}{2}+\frac{\alpha_s}{\pi}
\Bigl\{\frac{1}{72}(C_A(97+18L_m-18L_r)
\nonumber\\&&
~~~~~~~~~~
+4(9C_F-5T_Fn_l))
+\frac{1}{24}\beta_0(3L_r-8)\Bigr\}\biggr]
\,,
\eea
where $\delta_{\ell\ge 1}=1$ if ${\ell\ge 1}$ and zero otherwise.
The contributions of the annihilation diagrams are given by
\bea
&&
V_{\delta,a}=
\frac{\alpha_s^2}{m^2}C_F T_F\, \delta(\vec{r})
(4\log 2 - 4 )
\,,
\nonumber
\\ &&
V_{S,a}=
\frac{\alpha_s^2}{m^2}C_F T_F\, \vec{S}^2\, \delta(\vec{r})
(-2\log 2 + 2 )
\,.
\nonumber
\eea
The Fourier transforms are computed  using the
formula in App.~\ref{app1}, Eq.~(\ref{FTformula1})
and its derivatives with respect to $\vec{r}$.
``${\rm reg}[1/r^3]$'' in
$V_\delta$, $V_S$ can be identified with $1/r^3$ when applying to
$\ell > 0$ states but needs some modification when
computing matrix elements for the $\ell=0$ states.
We will come back to this point when explaining the
relevant computation
in Sec.~\ref{Subsec:NonCoulombCorr}.

\subsection{IR divergent potentials}
\label{Subsec:IRdivPot}

The IR divergent parts of the
Coulomb and non-Coulomb
potentials can be combined and are given
in momentum space as
\bea
\widetilde{V}^{pot}_{\rm div}
&=&
\widetilde{V}_{\rm C, div}+\widetilde{V}_{\rm NC, div}
\nonumber\\
&=&
-\frac{C_F\alpha_s}{6{\epsilon}}
\Biggl[
C_A^3\frac{\alpha_s^3}{q^2}+4(C_A^2+2C_AC_F)\frac{\pi\alpha_s^2}{mq}
+16\Bigl(C_F-\frac{C_A}{2}\Bigr)\frac{\alpha_s}{m^2}
+16C_A\frac{\alpha_s}{m^2}\frac{p^2+p^{\,\prime 2}}{2q^2}
\Biggr] .
\nonumber\\
\label{IRdivPot}
\eea
This corresponds to  Eq.~(4.59) of Ref. \cite{Beneke:2013jia}
(see also Ref. \cite{Beneke:2007pj}), in which $\delta V_{c.t.}$ is defined 
as a counter term and has the opposite sign compared to $\widetilde{V}^{pot}_{\rm div}$.

\section{NNNLO Corrections: Potential Region}
\label{Sec:PotRegion}

In this section we compute contributions to the NNNLO corrections to the
energy levels
which originate from the potential
region.
This corresponds to the pole positions 
of the singlet propagator as determined by the 
first diagram of Fig.~\ref{Diagrams-pNRQCD} or by the first term of
Eq.~(\ref{FullPropagatorS}).

\subsection{Potential perturbation}
\label{sec:Potential perturbation}
The leading-order Hamiltonian of the singlet
bound state is taken as that of the
pure Coulomb system:
\begin{eqnarray}
H_0 &=&\frac{\hat{\vec{p}}^{\, 2}}{m} -\frac{C_F\alpha_s}{r}.
\end{eqnarray}
All the other parts of $H_S$ in Eq.~(\ref{SingletHamiltonian})
are treated as perturbations.
We compute the perturbative expansion of the
singlet Green function, which 
can be  expressed as 
\begin{eqnarray}
G(\vec{r}, \vec{r}^{\,\prime}; E) &=& \langle \vec{r}| 
\frac{1}{(H_0+V_1+V_2+V_3+\cdots)-E-i 0 }  |\vec{r}^{\,\prime}\rangle
\nonumber
\\
&=& G^{(0)}+ \delta_1 G + \delta_2 G +\delta_3 G+\cdots 
,
\end{eqnarray}
where we denote by $V_i$ the $i$-th order potential\footnote{
The counting rule in the potential region is
$1/(mr)\sim p/m \sim \alpha_s$; see Eq.~(\ref{PotentialRegion}).
}.  The energy is measured from the threshold, $E=\sqrt{s}-2m$.
$G^{(0)}$ is the leading-order Green function given by
\begin{eqnarray}
G^{(0)} (\vec{r}, \vec{r}^{\,\prime}; E)= 
\langle \vec{r} |\,G^{(0)}(E) \, |\vec{r}^{\,\prime}\rangle ,
~~~~~~
G^{(0)}(E)=\frac{1}{H_0-E-i 0}
.
\end{eqnarray}
In the following we suppress 
$\vec{r}$, $\vec{r}^{\,\prime}$ and the infinitesimal 
imaginary part $i0$ of the Green function denominators.
Perturbative corrections to the Green function are
given by potential insertions:
\begin{eqnarray}
\delta_1 G(E) &=&
-\langle G^{(0)} \, V_1\, G^{(0)}\rangle,
\\
\delta_2 G(E) &=&
-\langle G^{(0)} \,V_2 \, G^{(0)}\rangle
+\langle G^{(0)} \,V_1 \, G^{(0)}  \, V_1\, G^{(0)}\rangle,
\\
\delta_3 G(E) &=& 
-\langle G^{(0)} \,V_3\, G^{(0)}\rangle
+\langle G^{(0)} \, V_1 \, G^{(0)}  \, V_2 \, G^{(0)}\rangle
\nonumber\\
&&
+\langle G^{(0)} \, V_2 \, G^{(0)}  \, V_1 \, G^{(0)}\rangle
-\langle G^{(0)}  \,  V_1  \,  G^{(0)}   \,  V_1 \,  G^{(0)} \, V_1 G^{(0)}\rangle.
\end{eqnarray} 
The Green function 
has  single poles corresponding to bound states
in the complex energy plane.
Suppose we are interested in the bound state, which is
specified at leading-order by $H_0\ket{n}=E_n^{(0)}\ket{n}$.
(In this subsection $n$ represents a set of quantum numbers
specifying a bound state, for simplicity of notations.)
The pole of the Green function for the corresponding bound
state can be written as
\begin{eqnarray}
G(\vec{r}, \vec{r}^{\,\prime}; E)
&\stackrel{E\to E_n}{=}&
\frac{
 F_n^{(0)}+F_n^{(1)} + F_n^{(2)} +F_n^{(3)} +\cdots 
}{\big [ E_n^{(0)}+E_n^{(1)}+E_n^{(2)}+E_n^{(3)}+\cdots \big] - E },
\label{PoleOfG}
\end{eqnarray}
where $E_n=\sum_i E_n^{(i)}$ and
$F_n=\sum_i F_n^{(i)}$
represent the perturbative expansions of the energy 
eigenvalue and residue (wave function) 
of the bound state, respectively. 
Hence, the perturbative expansion of Eq.~(\ref{PoleOfG}) is given by
\begin{eqnarray}
&&
\delta_1 G(E)
\stackrel{E\to E_n^{(0)}}{=}
\frac{F_n^{(1)} }{E_n^{(0)}-E}
-\frac{F_n^{(0)} E_n^{(1)}}{(E_n^{(0)}-E)^2},
\\
&&
\delta_2 G(E)  
\stackrel{E\to E_n^{(0)}}{=}
\frac{F_n^{(2)} }{E_n^{(0)}-E}
-\frac{F_n^{(1)} E_n^{(1)}+ F_n^{(0)} E_n^{(2)} }{(E_n^{(0)}-E)^2}
+\frac{F_n^{(0)} (E_n^{(1)})^2 }{(E_n^{(0)}-E)^3},
\\
&&
\delta_3 G(E)
\stackrel{E\to E_n^{(0)}}{=}
\frac{F_n^{(3)} }{E_n^{(0)}-E}
-\frac{ F_n^{(1)} E_n^{(2)} +F_n^{(2)} E_n^{(1)} +F_n^{(0)} E_n^{(3)} }{ \big(E_n^{(0)}-E \big)^2}
\nonumber\\
&& \hspace{2cm}
+\frac{ F_n^{(1)}\big( E_n^{(1)} \big)^2+2 F_n^{(0)} E_n^{(1)} E_n^{(2)} }{ \big(E_n^{(0)}-E \big)^3}
-\frac{ F_n^{(0)} \big( E_n^{(1)} \big)^3 }{ \big(E_n^{(0)}-E \big)^4}.
\end{eqnarray}
By comparison with the potential perturbation,
we obtain a master formula for the energy corrections 
up to NNNLO. 

It is convenient to define 
the reduced Green function and its first derivative with respect to  $E$
as follows.
\bea
&&
\overline{G}_n\equiv \lim_{E\to E_n^{(0)}} \biggl[{G}^{(0)}(E) -
\frac{|n\rangle \langle n| }{E_n^{(0)}-E}
\biggr]=
\sum_{m\neq n} \frac{|m\rangle \langle m| }{E_m^{(0)}-E_n^{(0)}} ,
\\ &&
\partial\overline{G}_n\equiv \lim_{E\to E_n^{(0)}}
\frac{\partial}{\partial E} \biggl[{G}^{(0)}(E) -
\frac{|n\rangle \langle n| }{E_n^{(0)}-E}
\biggr]=
\sum_{m\neq n} \frac{|m\rangle \langle m| }{(E_m^{(0)}-E_n^{(0)})^2} 
=\overline{G}_n^{\, 2} .
\eea
The energy and wave function corrections can be extracted from the double and single
poles, respectively.  
The first-order corrections can be extracted
from $\delta_1 G(E)$ and 
we obtain 
\begin{eqnarray}
E_n^{(1)}
&=&
\langle n| V_1 |n\rangle,
\\
F_n^{(1)}
&=&
-\bigg[ 
\langle \vec{r}| \overline{G}_n   V_1  |n\rangle \langle n|
\vec{r}^{\,\prime}\rangle
+\langle \vec{r}|n\rangle \langle n |  V_1 \overline{G}_n  |
\vec{r}^{\,\prime}\rangle
\bigg] .
\end{eqnarray}
The second-order corrections to the energy and wave function can be
extracted from the pole structure of $\delta_2G(E)$ and the results of 
the previous order. We obtain
\begin{eqnarray}
E_n^{(2)}
&=&
\langle n| V_2 |n\rangle
-\langle n| V_1\, \overline{G}_n \, V_1 | n\rangle,
\\
F_n^{(2)}
&=&
-\bigg[
\langle \vec{r}|\overline{G}_n  V_2  |n\rangle \langle n| \vec{r}^{\,\prime}\rangle
+\langle \vec{r} |n\rangle \langle n | V_2  \overline{G}_n | \vec{r}^{\,\prime}\rangle
\bigg]
\nonumber \\
&& 
+ \bigg[
\langle \vec{r}| n\rangle \langle n| V_1 \overline{G}_n   V_1  \overline{G}_n  |\vec{r}^{\,\prime}\rangle
+\langle \vec{r}| \overline{G}_n  V_1 |n\rangle \langle n| V_1 \overline{G}_n   |\vec{r}^{\,\prime}\rangle
\nonumber\\
&& 
\hspace{.5cm}
+ \langle \vec{r}| \overline{G}_n  V_1 \overline{G}_n   V_1 |n \rangle  \langle  n| \vec{r}^{\,\prime} \rangle
\bigg]
- \bigg[
\langle \vec{r} |n\rangle \langle n| V_1 |n\rangle \langle n|  V_1\partial\overline{G}_n  |\vec{r}^{\,\prime}\rangle
\nonumber\\
&& 
\hspace{.5cm}
+ \langle \vec{r}| n\rangle \langle n|  V_1  \partial\overline{G}_n   V_1 |n\rangle \langle n|\vec{r}^{\,\prime}\rangle
+ \langle \vec{r} | \partial\overline{G}_n  V_1  | n \rangle \langle n| V_1 |n \rangle \langle n |\vec{r}^{\,\prime}\rangle
\bigg].
\end{eqnarray}
The energy correction at third order is given by 
\begin{eqnarray}
E_n^{(3)}
&=&
\langle n| V_3 | n\rangle
-2 \langle n| V_1\, \overline{G}_n  \, V_2 | n\rangle
\nonumber\\
&&
+\, \langle n| V_1  \overline{G}_n  V_1  \overline{G}_n  V_1  |n \rangle
-E_n^{(1)} \langle n| V_1 \partial \overline{G}_n  V_1 |n\rangle .
\label{MasterFormulaE3}
\end{eqnarray}

Thus, for the third order corrections to the energy levels, we need to compute 
a single insertion $\langle n| V_3 | n\rangle$, 
double insertions $ \langle n| V_1\, \overline{G}_n  \, V_2 | n\rangle, 
\langle n| V_1 \partial \overline{G}_n  V_1 |n\rangle$
and a triple insertion $\langle n| V_1  \overline{G}_n  V_1  \overline{G}_n  V_1  |n \rangle$.
The details of methods to compute all the insertions are explained in the following subsections.

\subsection{Wave functions and Green function of Coulomb system}
\label{sec:Coulomb-system}

To apply the above master formula we need explicit
representations for the wave functions and the
Green function of the leading-order
Coulomb system.
These are given as follows.

Let us define the relevant variables as
\begin{eqnarray}
m_r &=& \frac{m}{2} ,
\\
a_s^{-1} &=& m_r \alpha_s C_F  ,
\\
a_o^{-1} &=& m_r \alpha_s \left(\frac{C_A}{2}-C_F\right)  ,
\\
\lambda &=& \frac{m_r \alpha_s C_F }{ \sqrt{-2m_rE}} .
\end{eqnarray}
The radial part of the  wave function
for the color-singlet bound state is give by
\begin{eqnarray}
&&
R_{n\ell}(r)=
N_{n\ell} \,z_n^{\, \ell} 
e^{-\frac{1}{2} z_n } 
L_{n-\ell-1}^{2\ell+1}\left(z_n\right),
\\ &&
z_n=\frac{2r}{n a_s},
~~~
N_{n\ell} =
\left(\frac{2}{n a_s}\right)^{\frac{3}{2}}
\sqrt{\frac{(n-\ell-1)!}{2n\, (n+\ell)!}},
\label{SingletRadialWaveFn}
\end{eqnarray}
which is normalized as
$\int_0^{\infty} dr\, r^2 \, R_{n\ell}(r) R_{n' \ell}(r) = \delta_{n n'}$.
The definition of the Laguerre polynomial
$L_{m}^{a}(z )$ is given in App.~\ref{app:CoulombME},
Eq.~(\ref{GeneFnLaguerre}).

The octet wave function is given by 
\begin{eqnarray}
R^{(o)}_{k\ell} (r)&=&  
A_{k\ell} \, (2kr)^{\ell} \,e^{-i kr}
{_1 F_1}\bigg(-\frac{i}{ka_o}+\ell+1; 2\ell+2; 2ikr\bigg),
\\
A_{k\ell}
&=&
\frac{1}{a_o (2\ell+1)!}
\sqrt{ \frac{8\pi k a_o }{e^{2\pi/(ka_o)}-1}}\,
\prod_{s=1}^\ell \sqrt{s^2+\frac{1}{k^2 a_o^2}},
\end{eqnarray}
which is normalized as
$\int_0^\infty dr\, r^2 R_{k\ell}^{(o)}(r) R_{k'\ell}^{(o)}(r) =2\pi\delta(k-k')$.
The confluent hypergeometric function is defined as
$
_1F_1(a;c;z)=\sum_{n=0}^\infty \frac{a(a+1)\cdots(a+n-1)}
{c(c+1)\cdots(c+n-1)}\frac{z^n}{n!}
$.

We use the following representation of the
Coulomb Green function for our computation:
\begin{eqnarray}
G(\vec{r}_1, \vec{r}_2; E) 
&=&
\sum_{\ell =0}^\infty
\frac{2\ell+1}{4\pi} P_\ell ({\hat{r}_1 \cdot \hat{r}_2} )\, G_{\ell}(r_1, r_2; E) ,
\end{eqnarray}
where $P_\ell(x)=\frac{1}{2^\ell \ell!}\frac{d^\ell}{dx^\ell}
(x^2-1)^\ell$
 denotes the Legendre polynomial, and
$\hat{r}_i=\vec{r}_i/r_i$ represents the unit vector in
the direction of $\vec{r}_i$.
The Green function for the partial wave $\ell$ is 
given by an infinite sum\footnote{
This representation of $G$ includes contributions of not
only the bound states at $E<0$ but also those of the continuum states
at $E>0$, although this may not be obvious from the summation formula.
}
\begin{eqnarray}
G_{\ell}(r_1, r_2; E)&=& 
\sum_{\nu=\ell+1}^\infty  
G_{\nu \ell}(r_1,r_2; E),
\\
G_{\nu\ell}(r_1, r_2; E)
&=&
m_r a_s^2\,
\left( \frac{\nu^4}{\lambda}\right)
\frac{R_{\nu\ell}(z_{\lambda 1}) R_{\nu\ell}(z_{\lambda 2} )}{\nu-\lambda}
,
\hspace{1cm}
\biggl(z_{\lambda i} = \frac{2r_i}{\lambda a_s}\biggr) .
\end{eqnarray}
where $R_{\nu\ell}$
is the singlet bound-state wave function.

The reduced Green function $\overline{G}_{n\ell}$ for the
partial wave $\ell$ can
be computed as follows.
Let
\begin{eqnarray}
{\cal G}_{n\ell}(r_1, r_2; E) &\equiv& 
\frac{R_{n\ell}(z_{n1}) R_{n\ell} (z_{n2}) }{E_n^{(0)}-E}
\nonumber \\
&=&
m_r a_s^2\, 
\left( \frac{2n^2\lambda^2}{n+\lambda}\right)
\frac{R_{n\ell}(z_{n 1}) R_{n\ell}(z_{n 2} )}{n-\lambda},
\hspace{1cm}
\biggl(z_{ni}= \frac{2r_i}{na_s}\biggr)
\nonumber.
\end{eqnarray}
Then
\begin{eqnarray}
\overline{G}_{n\ell}(r_1,r_2) &\equiv&
\lim_{E\to E_n^{(0)}} \left\{ 
\sum_{\nu=\ell+1}^\infty G_{\nu\ell}(r_1,r_2;E) 
- {\cal G}_{n\ell}(r_1,r_2; E)
\right\}
\nonumber \\
&=& \sum_{\nu=\ell+1}^\infty 
\overline{G}_{\nu, n\ell}(r_1,r_2), 
\label{eq:reduced-Green-function}
\eea
with
\bea
\overline{G}_{\nu, n\ell} (r_1 ,r_2) 
&=&
\left\{
\begin{array}{ll}
\displaystyle
\vspace*{3mm}
m_r a_s^2 \left( \frac{\nu^4}{n}\right)
\frac{R_{\nu\ell}(z_{n1}) R_{\nu\ell}(z_{n2}) }{\nu-n}
&
~~(\nu\neq n)
\\
\displaystyle
m_r a_s^2 n^2 
\left\{\frac{5}{2} + z_{n1}\frac{d}{d z_{n1}}+ z_{n2}\frac{d}{d z_{n2}} \right\} 
R_{n\ell}(z_{n1})R_{n\ell}(z_{n2})
&
~~(\nu= n)
\end{array}
\right. .
\nonumber\\
\end{eqnarray}
The first derivative of the 
radial wave function can be rewritten in terms of the same 
radial wave function and 
the one with the index $\beta(=n-\ell-1)$ lowered by one:
\begin{eqnarray}
z_n \frac{dR_{n\ell}(z_n) }{dz_n} 
&=&
\bigg[
\left\{ (n-1) -\frac{z_n}{2} - (n+\ell ) \hat{d}_{\beta} \right\} 
R_{n\ell}(z_{n}) 
\bigg]_{\beta=n-\ell-1},
\\
R_{n\ell}(z_n) 
&=&
N_{n\ell} \, z_n^{\ell} e^{-\frac{1}{2} z_n} L_\beta^{(2\ell+1)}(z_n) 
\hspace{1cm} \mbox{with}~~~ \beta=n-\ell-1 ,
\label{index-lowering}
\end{eqnarray}
where $\hat{d}_\beta$ is the index-lowering operator for $\beta$ [subscript of $L^{(2\ell+1)}_\beta(z_n)$].

\subsection{Coulomb corrections}

We compute the corrections which originate purely from
the renormalized Coulomb potential
Eq.~(\ref{V_{c,ren}(r)}) in the potential perturbation of
Sec.~\ref{sec:Potential perturbation}.
Using the reduced Green function in 
Sec.~\ref{sec:Coulomb-system} and
formulas in App.~\ref{app:CoulombME},
it is cumbersome but straightforward to
obtain the Coulomb corrections 
expressed as combinations of
infinite sums.
Then using techniques similar to
that of Sec.~\ref{sec:NNLO}, we can express
the corrections with finite sums.

More explicitly, we proceed as follows.
We compute $E_n^{(3)}$ by Eq.~(\ref{MasterFormulaE3}).
$V_i$ is identified with the order $\alpha_s^{i+1}$ term of
Eq.~(\ref{V_{c,ren}(r)}).
We use the wave function in Eq.~(\ref{SingletRadialWaveFn}) and
the reduced Green function
Eqs.~(\ref{eq:reduced-Green-function})--(\ref{index-lowering})
to express the matrix elements.
Integrals over angular variables can be performed using
the relation
\bea
\int d\Omega_{\hat{s}}\, P_\ell(\hat{r}\cdot\hat{s})\,
P_\ell(\hat{r}'\cdot\hat{s})
=\frac{4\pi}{2\ell+1}\,P_\ell(\hat{r}\cdot\hat{r}') .
\eea
Integrals over the radial variables 
can be performed using the formulas
Eqs.~(\ref{DiagLaguerreFormula1})--(\ref{DiagLaguerreFormula-1})
and 
Eqs.~(\ref{OffdiagLaguerreFormula1})--(\ref{OffdiagLaguerreFormula-1}).
All the infinite sums that appear can be converted to 
combinations of transcendental numbers and finite sums
using the program \texttt{Wa} and the technique described
in Sec.~\ref{sec:NNLO}.

\subsection{Non-Coulomb corrections}
\label{Subsec:NonCoulombCorr}

We compute the NNNLO corrections
which involve the non-Coulomb potentials.
Since there are no NLO potentials in
the non-Coulomb potentials,
there are only double insertions and single insertion
in the potential perturbation.

\subsection*{Double insertion: spin-dependent part}

We first consider the spin-dependent part of 
the double insertion of potentials.
This contribution is given by
\bea
E_{\rm 2,a}^{\rm NC}=2
\bra{n\ell sj}V_1 \cdot \overline{G}_{n\ell} \cdot U_2
\ket{n\ell sj}
\eea
where
\bea
V_1=-\frac{C_F\alpha_s}{r}\,\frac{\alpha_s}{4\pi}
\Bigl[a_0\beta_0L_r+a_1\Bigr]
\eea
denotes the NLO Coulomb potential, and
\bea
U_2=
\frac{\pi C_F \alpha_s}{m^2}
\bigg\{ \frac{4}{3}\,\delta(\vec{r})\, \vec{S}^2
+\frac{3 }{2\pi r^3} \vec{L}\cdot \vec{S}
+\frac{1}{2\pi r^3}
 \biggl(3 \frac{(\vec{r}\cdot \vec{S})^2}{r^2} - \vec{S}^2  \biggr)
\bigg\}
\eea
represents the spin-dependent part of the NNLO non-Coulomb
potentials.

In computing $E_{\rm 2,a}^{\rm NC}$ we can replace $U_2$
by
\bea
U'_2=
\frac{\pi C_F \alpha_s}{m^2}
\bigg\{ \frac{4}{3}\,\delta(\vec{r})\, \mathbb{S}^2
+\frac{3 }{2\pi r^3} X_{LS}
+\frac{1}{2\pi r^3}
 D_{S}
\bigg\}
\eea
where
\bea
&&
\mathbb{S}^2\equiv \left< \vec{S}^2 \right> =s(s+1) ,
\\&&
X_{LS} \equiv
\left< \vec{L}\cdot \vec{S} \right>
= \frac{1}{2}\,
\left[ j(j+1)-\ell(\ell+1)-s(s+1) \right] ,
\\ &&
D_{S} \equiv
\biggl< 
3 \frac{(\vec{r}\cdot \vec{S})^2}{r^2} - \vec{S}^2 
\biggr>
=
\frac{
2 \ell (\ell+1) s (s+1) - 3 X_{LS} - 6 X_{LS}^2
}{
(2\ell-1)(2\ell+3)
}.
\label{DefDS}
\eea
The expectation value $\langle \cdots \rangle$
is taken with respect to the $(\ell,s,j)$ state.
The derivation of $D_S$ is given in App.~\ref{App:DS}.
The above replacement is justified, since 
$V_1$ and $\overline{G}_{n\ell}$ do not change
$\ket{\ell sj}$, hence only matrix elements proportional to
$\bra{\ell sj}U_2\ket{\ell sj}=U_2'$ appear.

One can compute $E_{\rm 2,a}^{\rm NC}$
easily similarly to the Coulomb corrections.
One needs to compute the corrections for
the
$\ell=0$ and $\ell> 0$ cases separately,
due to existence of $\delta(\vec{r})$ and $1/r^3$.
In particular, in the case $\ell=0$ one should set $X_{LS}=D_S=0$
before evaluating the matrix elements of $1/r^3$.
(See explanation for the single insertion below.)

\subsection*{Double insertion: spin-independent part}

Next we consider the spin-independent part of 
the double insertion of potentials.
This contribution is given by
\bea
E_{\rm 2,b}^{\rm NC}=2
\bra{n\ell sj}V_1 \cdot \overline{G}_{n\ell} \cdot U_1
\ket{n\ell sj}
\eea
where
\bea
U_1=
 -\frac{\hat{\vec{p}}^{\,4}}{4m^3}-
 \frac{C_F \alpha_s}{2m^2} \bigg\{\hat{\vec{p}}^{\,2},\frac{1}{ r}\bigg\}  
 +
 \frac{C_F(C_F-2C_A)\alpha_s^2}{4m r^2} 
\eea
represents the spin-independent part of the NNLO non-Coulomb
potentials.

$E_{\rm 2,b}^{\rm NC}$ can be computed easily by eliminating
the $\hat{\vec{p}}$ operator in the following way.
\begin{eqnarray}
&& - \frac{\hat{\vec{p}}^{\,4}}{4m^3} 
=-\frac{1}{4} ( U_A - U_B -U_C+U_D) ,
\\
&& 
 \frac{1}{2m^2} \bigg\{\hat{\vec{p}}^{\,2}, V_0  \bigg\}
=\frac{1}{2}( U_B+U_C-2U_D)
\end{eqnarray}
with $V_0=-C_F\alpha_s/r$  and
\begin{eqnarray}
U_A =\frac{H_0^2}{m} ,\hspace{.5cm}
U_B = \frac{\{H_0-E_n^{(0)}, V_0 \}}{m}  ,\hspace{.5cm}
U_C= \frac{2E_n^{(0)} V_0}{m} ,\hspace{.5cm}
U_D= \frac{V_0^2}{m}.
\end{eqnarray}
Matrix elements which are needed to evaluate the double insertions of 
the NLO Coulomb and $U_1$
are given as
\begin{eqnarray}
&&
\langle n| V_1 \overline{G}_n U_A |n\rangle
=0 ,
\\
&&
\langle n| V_1 \overline{G}_n U_B |n\rangle
=\frac{1}{m} \bigg\{ \langle n| V_1 V_0  |n\rangle-\langle n| V_1 |n\rangle\langle n| V_0 |n\rangle\bigg\} ,
\\
&&
\langle n| V_1 \overline{G}_n U_C |n\rangle
=\frac{2E_n^{(0)}}{m} \langle n| V_1 \overline{G}_n  V_0|n\rangle ,
\\
&&
\langle n| V_1 \overline{G}_n U_D |n\rangle
=
\langle n| V_1 \overline{G}_n \, \frac{V_0^{\, 2}}{m} \, |n\rangle .
\end{eqnarray}

\subsection*{Single insertion}

Finally we compute the single insertion of the
NNNLO non-Coulomb potentials
in potential perturbation:
\bea
E_{\rm 1}^{\rm NC}=
\bra{n\ell sj}V_{\rm 3,NC}\ket{n\ell sj} .
\eea
We use the renormalized non-Coulomb potentials in
coordinate space
given in Eq.~(\ref{RenormalizedNonCoulPot}).
We take only the NNNLO part
as $V_{\rm 3,NC}$, namely those potentials which 
are not included in $U_1+U_2$.

In computing the above matrix element, one needs to be
careful in evaluating the contribution of $V_\delta$.
The operators $\delta(\vec{r})$ and ${\rm reg}[1/r^3]$ are 
singular when their matrix elements are computed
with respect to $\ell=0$ states.
This is an artifact of working in coordinate space, since
no singularities arise if the same matrix elements
are evaluated in momentum space.
Nevertheless, it is advantageous to work in coordinate
space for obtaining a general formula for arbitrary
$(n,\ell,s,j)$.
Therefore, we compute in the following way.

$\delta(\vec{r})$ and ${\rm reg}[1/r^3]$, respectively, stem from
the Fourier transforms of 1 and $L_q=\log(\mu^2/q^2)$.
There are no singularities if we define them
as the $u\to 0$ limits of $(\mu^2/q^2)^u$ and 
$\partial_u [(\mu^2/q^2)^u]$, respectively, in momentum space. 
Hence, we first take the Fourier transform
of $(\mu^2/q^2)^u$ and $\partial_u [(\mu^2/q^2)^u]$
using the formula Eq.~(\ref{FTformula1}),
then evaluate the matrix elements in coordinate space 
using the formulas in App.~\ref{app:CoulombME}.
We take the limit $u\to 0$ in the end  and obtain
the matrix elements.
In this way we confirm that the operator $\delta(\vec{r})$
can be handled in the standard way.
On the other hand, one finds that the matrix elements of
the operator ${\rm reg}[1/r^3]$
should be evaluated as
\bea
&&
\biggl\langle n,\ell=0 \biggl|{\rm reg}\biggl[ \frac{1}{r^3}\biggr] 
\biggr| n,\ell=0
\biggr\rangle 
=\frac{1}{4}\biggl(\frac{C_F\alpha_sm}{n}\biggr)^3
\nonumber\\&&
~~~~~~~~~~~~~~~~~~~~~~~~~~~~~~~~~~~~
\times
\biggl\{ 2\log\biggl(\frac{n\mu}{C_F\alpha_sm}\biggr)
-2S_1(n)+\frac{1}{n}-1\biggr\}
\eea
for the $\ell=0$ states, whereas ${\rm reg}[1/r^3]$
can be identified with $1/r^3$ for the matrix elements
with respect to $\ell>0$ states.

\subsection{IR divergent contributions}
\label{IRdivEnergy}

The IR divergent contributions which appear first at NNNLO
is given by the single insertion of the divergent potentials
Eq.~(\ref{IRdivPot}):
\bea
E^{pot}_{\rm div} =
\bra{n\ell sj}{V}^{pot}_{\rm div}\ket{n\ell sj} .
\eea
As we will see in the next section, this entire contribution is canceled by
the UV divergent part of the corrections $\langle n\ell sj| H_{\rm div}^{us}|n\ell sj \rangle$ from the ultra-soft region.
Therefore, we do not evaluate explicitly the matrix element for the divergent contributions.

\section{NNNLO Corrections: Ultra-soft Region}

In this section we compute contributions to the NNNLO corrections to the
energy levels
which originate from both ultra-soft and potential
regions.
This corresponds to the self-energy 
of the singlet propagator in the 
second diagram of Fig.~\ref{Diagrams-pNRQCD} or in the second term of
Eq.~(\ref{FullPropagatorS}).

\subsection{Separation of local and non-local parts}

The one-loop self-energy of the singlet field $S$
is given by
\bea
E^{us}_{n\ell}=-ig^2 \bar{\mu}^{2\epsilon}
\frac{T_F}{N_C}\int_0^\infty dt 
\left\langle \vec{r}\cdot\vec{E}^a(t,\vec{0})\,
\exp\left[ -i(H_O^{(d)} - E_{n,C}^{(d)})t \right]
\vec{r}\cdot\vec{E}^a(0,\vec{0})
\right\rangle_{n\ell}
.
\label{DefEUS}
\eea
We set $U_O(t)^{ab}\approx e^{-iH_Ot}\delta^{ab}\theta(t)$ in 
Eq.~(\ref{FullPropagatorS}), which corresponds to
replacing $D_t$ in Eq.~(\ref{DefOctetProp}) by $\partial_t$,
where
the omitted terms are suppressed by additional powers
of $\alpha_s$.
$H_O^{(d)}$ denotes the Hamiltonian for the octet $Q\bar{Q}$ states in
$d$ dimensions, while
$E_{n,C}^{(d)}=-\frac{C_F^2\alpha_s^2}{4n^2}m + {\cal O}(\epsilon)$ 
denotes the (leading-order) energy eigenvalue of the singlet
Hamiltonian in $d$ dimensions, $H_S^{(d)}$:
\bea
&&
H_S^{(d)}=\frac{\hat{\vec{p}}^{\,2}}{m}+V_S^{(d)}(r)
\, ,~~~
H_O^{(d)}=\frac{\hat{\vec{p}}^{\,2}}{m}+V_O^{(d)}(r) \, ,
\\
&&
V_S^{(d)}(r)=-C_F\frac{\alpha_s}{r}
\left(\bar{\mu} r\right)^{2\epsilon} A(\epsilon)
\, ,~~~
V_O^{(d)}(r)=\left(\frac{C_A}{2}-C_F\right)\frac{\alpha_s}{r}
\left(\bar{\mu} r\right)^{2\epsilon} A(\epsilon) \, ,
\\ &&
A(\epsilon)=\frac{\Gamma({\scriptstyle\frac{1}{2}}-\epsilon)}
{\pi^{\frac{1}{2}-\epsilon}}
.
\eea
In Eq.~(\ref{DefEUS}),
$\langle \cdots \rangle_{n\ell}$ denotes the expectation value
taken with respect to the energy eigenstate of $H_S^{(d)}$
specified by the quantum numbers $(n,\ell,s,j,j_z)$
(the expectation value depends only on $n,\ell$).
Only the leading-order terms of $H_S$ and $H_O$ are needed
in our computation, whereas we need to keep 
$\epsilon=(3-d)/2$ non-zero until we extract the UV divergence
of $E^{us}_{n\ell}$ explicitly.

The gluon emitted and reabsorbed by $Q\bar{Q}$ after the
time interval $t$ is in the ultra-soft region.
On the other hand, the propagator of the octet field
$e^{-iH_Ot}$
contains multiple exchanges of Coulomb gluons
which are in the potential region.
Therefore the whole correction consists of a combination
of contributions from the ultra-soft and potential regions.

We separate $E^{us}_{n\ell}$ into two parts:
the part given by an
expectation value of a Hamiltonian (local in time) and the part
given by a term known as the ``QCD Bethe logarithm'' (non-local
in time).
This is achieved in the following way.
The correlation function of the color electric
field (non-local in time and local in space) is evaluated
in the lowest order in expansion in $\alpha_s$ as
\bea
&&
\left\langle E^{ia}(t,\vec{0})E^{ja}(0,\vec{0})
\right\rangle
=-i\delta^{aa}\int \frac{d^Dk}{(2\pi)^D}\,
\frac{e^{ik_0t}}{k^2+i0}\,\left(k^ik^j-k_0^2\delta^{ij}\right)
~+~{\cal O}(\alpha_s)
\nonumber\\
&&~~~~~~~~~~~~~~~ ~~~~~~~~~~~
=\frac{C_AC_F}{2T_F}\,\frac{d-1}{d}\,\delta^{ij}
\int \frac{d^{d}\vec{k}}{(2\pi)^{d}}\, |\vec{k}| \,
e^{-i|\vec{k}|t}
~+~{\cal O}(\alpha_s) .
\eea
After integrating over $t$ and using
\bea
&&
\int \frac{d^{d}\vec{k}}{(2\pi)^{d}}\, 
\frac{|\vec{k}| }{|\vec{k}| +\kappa}
= C(d)\, \kappa^{d} 
~~~~;~~~~
C(d)=2^{1-d}\pi^{-d/2}\,
\frac{\Gamma(1+d)\Gamma(-d)}{\Gamma(d/2)} ,
\eea
we obtain
\bea
&&
E^{us}_{n\ell}=\frac{1}{2}C_F g^2 \bar{\mu}^{2\epsilon}
\, \frac{1-d}{d}\, C(d)\,
\left\langle r^i \left(H_O^{(d)} - E_{n,C}^{(d)}\right)^d
r^i \right\rangle_{n\ell}
.
\label{EUS2}
\eea
Note that $C(d)$ is order $1/\epsilon$.

We may expand
$(H_O-E_{n,C})^{3-2\epsilon}\approx (H_O-E_{n,C})^{3}
[1-2\epsilon\log(H_O-E_{n,C})]$ and write
\bea
&&
\bar{\mu}^{2\epsilon}
\left\langle r^i \left(H_O^{(d)} - E_{n,C}^{(d)}\right)^d
r^i \right\rangle_{n\ell}
= \left\langle
X - 2\epsilon\, r^i \left(H_O^{(d)} - E_{n,C}^{(d)}\right)^3
\log \left(\frac{H_O^{(d)} - E_{n,C}^{(d)}}{\bar{\mu}}\right)
r^i \right\rangle_{n\ell}
\nonumber \\
&&
~~~~~~~~~~~~~~~~~~~~
~~~~~~~~~~~~~~~~~
~~~~~~~~~~~
+ {\cal O}(\epsilon^2)
\eea
with
\bea
X=r^i \bigl(H_O^{(d)}\bigr)^3 r^i- \frac{3}{2}
\left\{ H_S^{(d)}, r^i \bigl(H_O^{(d)}\bigr)^2 r^i \right\}
+ \frac{3}{2}
\left\{ \bigl(H_S^{(d)}\bigr)^2, r^i H_O^{(d)} r^i \right\}
- \frac{1}{2}
\left\{ \bigl(H_S^{(d)}\bigr)^3, \vec{r}^{\, 2} \right\}
,
\nonumber\\
\label{defX}
\eea
where we have replaced $E_{n,C}^{(d)}$ by the singlet Hamiltonian
inside the expectation value,
taking into account ordering of the operators.
We rewrite $X$ by
moving all the momentum operators $\hat{\vec{p}}=-i\vec{\partial}_r$ 
to the right of the
coordinate operators $\vec{r}$ and $r=|\vec{r}|$ using
commutation relations, and then we Fourier transform the result.
After expanding in $\epsilon$ and using the equation of motion,
we obtain (see App.~\ref{appHUS} for details)
\bea
&&
E^{us}_{n\ell}=
\left\langle H^{us} \right\rangle_{n\ell}
-\frac{2C_F\alpha_s}{3\pi}\left\langle
r^i \left(H_O^{(d)} - E_{n,C}^{(d)}\right)^3
\log \left(\frac{H_O^{(d)} - E_{n,C}^{(d)}}{{\mu}}\right)
r^i \right\rangle_{n\ell}
.
\label{EUS3}
\eea
This algebraic derivation for the ultra-soft Hamiltonian  is performed
in $d$ dimensions and agrees with the effective Hamiltonian  $(H^{us}+\delta H^{us})$  
of  Ref. \cite{Kniehl:2002br} derived from  the Feynman diagram method 
in momentum space. They are given by 
\bea
&&
\widetilde{H}^{us}=\widetilde{H}^{us}_{\rm div}+\widetilde{H}^{us}_{\rm ren}
,
\\
&&
\widetilde{H}^{us}_{\rm div}=-\widetilde{V}^{pot}_{\rm div}
\label{HUSdiv}
,
\\
&&
\widetilde{H}^{us}_{\rm ren}=
\frac{8C_F(C_A-2C_F)\alpha_s^2}{3m^2}\left(
\log 2-\frac{5}{6}\right)-\frac{16C_FC_A\alpha_s^2}{3m^2}\,
\frac{p^2+p'^2}{2q^2}\left(
\log 2-\frac{5}{6}\right)
\nonumber\\&&
~~~~~~
+
\frac{\pi C_FC_A\alpha_s^3}{9mq}\left\{
6(C_A+2C_F)\log\Bigl(\frac{{\mu}^2}{q^2}\Bigr)-4C_F+13C_A\right\}
\nonumber\\&&
~~~~~~
+
\frac{C_FC_A^3\alpha_s^4}{3q^2}
\left\{\log\Bigl(\frac{{\mu}^2}{q^2}\Bigr)-\log 2+\frac{5}{6}
\right\}
.
\label{HUSren}
\eea
Thus, the UV divergent part of $H^{us}$ is canceled by
the IR divergent part of the singlet potentials;\footnote{
This implies that the EFT is independent of the factorization
scale between the potential and ultra-soft regions.
} 
see Secs.~\ref{Subsec:IRdivPot}
and \ref{IRdivEnergy}.
The renormalized Hamiltonian in coordinate space reads
\bea
&&
H^{us}_{\rm ren}=\frac{C_FC_A\alpha_s^3}{3\pi mr^2}
\biggl[\frac{1}{6}(13C_A-4C_F)+(C_A+2C_F)L_r\biggr]
\nonumber\\&&~~~~~~~~
+\frac{2C_FC_A\alpha_s^2}{3\pi m^2}\left(\frac{5}{6}-\log 2\right)
\biggl\{ \hat{\vec{p}}^{\,2} ,\frac{1}{r} \biggr\}
+\frac{C_FC_A^3\alpha_s^4}{12\pi r}
\biggl( L_r+\frac{5}{6}-\log 2 \biggr)
\nonumber\\&&~~~~~~~~
-\frac{8C_F\alpha_s^2}{3m^2}\delta(\vec{r})
(C_A-2C_F)\left(\frac{5}{6}-\log 2\right)
.
\eea

After adding the contributions from the counter terms,
the ultra-soft correction to the bound state energy becomes
finite and
is given by 
\begin{eqnarray}
E^{us}_{n\ell,{\rm ren}}
&=&
E^{us}_{n\ell}+E^{pot}_{\rm div}
\nonumber\\
&=&
\langle {H}_{\rm ren}^{us} \rangle_{n\ell}
-\frac{2C_F\alpha_s}{3\pi}
\log \Bigl(\frac{|E_1^C|}{{\mu}}\Bigr)\,
\left\langle
r^i \left(H_O^{(3)} - E_n^C\right)^3
r^i \right\rangle_{n\ell}
\nonumber\\ &&
+\frac{2(C_F\alpha_s)^3}{3\pi}|E_n^C|\,L_{\rm Bethe}(n,\ell)
,
\label{EUSren}
\eea
where $E_n^C=E_{n,C}^{(3)}=-\frac{C_F^2\alpha_s^2}{4n^2}m$.
We have split the logarithmic term into
the $\mu$ dependent part and $\mu$ independent part,
where the QCD Bethe logarithm is defined as
\bea
L_{\rm Bethe}(n,\ell)=
-\frac{1}{(C_F\alpha_s)^2|E_n^C|}
\left\langle
r^i \left(H_O^{(3)} - E_n^C\right)^3
\log \left(\frac{H_O^{(3)} - E_n^C}{|E_1^C|}\right)
r^i \right\rangle_{n\ell}
.
\eea
All the other parts can be evaluated analytically.
In particular, the expectation value in 
the second term of Eq.~(\ref{EUSren})
can be written as
\bea
&&
\left\langle
r^i \left(H_O^{(3)} - E_n^C\right)^3
r^i \right\rangle_{n\ell}
\nonumber\\&&
=
\left\langle
\frac{C_A^3\alpha_s^3}{8r}+\frac{C_A(C_A+2C_F)\alpha_s^2}{mr^2}
+\frac{C_A\alpha_s}{m^2}\left\{\hat{\vec{p}}^{\, 2},\frac{1}{r}
\right\}
+\frac{4\pi(2C_F-C_A)\alpha_s}{m^2}\,\delta(\vec{r})
\right\rangle_{n\ell} ,
\nonumber\\
\eea
which is related to the UV divergence.

\subsection{One-parameter integral form of QCD Bethe logarithm}
\label{sec:BetheLog}
The QCD Bethe logarithm for the S-wave states was  first computed in Ref. \cite{Kniehl:1999ud}.
We derive the one-parameter integral form for the Bethe logarithm 
for a general quantum number in the following way. 
We insert the completeness relation for the octet states:
\bea
&&
-{(C_F\alpha_s)^2|E_n^C|}L_{\rm Bethe}(n,\ell)=
\sum_{\ell', \ell_z'}
\int_0^\infty 
\frac{d{k}}{2\pi}
\left(\frac{{k}^2}{m} - E_n^{C}\right)^3
\nonumber\\
&&
~~~~~~~~~~~~~~~~~~
~~~~~~~~~~~~~~~~~~
\times
\log\left( {\frac{{{k}^2}/{m}-E_n^C}{\mu}} \right)
\,
|\langle \psi_{k\ell' \ell_z'}^{(o)}| \, {r}^i \,|\psi_{n\ell \ell_z}^{(s)}\rangle|^2 .
\end{eqnarray}
Decomposing the radial and angular parts of the wave function
we obtain (see App.~\ref{AppAngAve} for derivation)
\begin{eqnarray}
\sum_{\ell' \ell_z'}
|\langle \psi_{k\ell' \ell_z'}^{(o)}| \, {r}^i \,|\psi_{n\ell \ell_z}^{(s)}\rangle|^2
&=&
\sum_{\ell'}
|\langle R_{k\ell'}^{(o)}| \, {r} \,|R_{n\ell }^{(s)}\rangle|^2\,
\frac{{\rm max}(\ell,\ell') }{2\ell+1} 
\left( \delta_{\ell-1,\ell'} +\delta_{\ell+1,\ell'}\right)
.
\label{AngAve}
\end{eqnarray}
Using the wave functions in Sec.~\ref{sec:Coulomb-system},
the radial part can be written as
\begin{eqnarray}
\langle R_{k\ell'}^{(o)}| r | R_{nl}^{(s)}\rangle
&=&
\frac{96 \sqrt{2\pi}(na_0)^{\frac{3}{2}}
      }{(2\ell+1)! (2\ell'+1)!}
\sqrt{\frac{(n+\ell)!}{n\, (n-\ell-1)!} \, 
\frac{\hat{k}  }{(e^{2\pi/\hat{k}}-1) }
\prod_{s=1}^{\ell'} \left(\frac{1}{\hat{k}^2} + s^2 \right)
}
\nonumber\\
&&\times
\frac{ \, \hat{k}^{\ell'} \,\rho_n^{\ell'+1}  e^{\frac{2}{\hat{k}} \arctan(\hat{k}\rho_n)} 
      }{ (1+\hat{k}^2 \rho_n^2 )^{(\ell+\ell'+5)/2}} \, X_{n\ell}^{\ell'}
\end{eqnarray}
with
\begin{eqnarray}
&&
k=\frac{\hat{k}}{a_o},~~
r=\frac{n a_s}{2} z, ~~
\rho_n = \frac{a_s}{a_o }n 
.
\eea
The matrix element $X_{n\ell}^{\ell'}$ is defined as 
\begin{eqnarray}
\frac{ 384 e^{\frac{2}{\hat{k}} \arctan(\hat{k}\rho_n)} 
      }{ (1+\hat{k}^2 \rho_n^2 )^{(\ell+\ell'+5)/2}}
 X_{n\ell}^{\ell'}
&\equiv &
\int_0^\infty dz \, z^{\ell+\ell'+3}
e^{-\frac{1}{2}(1+i \hat{k}\rho_n) z}
\nonumber\\
&&\times
 {_1 F_1}(-n+\ell+1;\, 2\ell+2;\, z)
 \nonumber\\
 &&\times
 {_1 F_1}(-i/\hat{k}+\ell'+1;\, 2\ell'+2; \, i\hat{k}\rho_n z).
\end{eqnarray}

Changing the variable to $x=1/\hat{k}$, we find
\bea
L_{\rm Bethe}(n,\ell)=\int_0^\infty dx
\sum_{
\mathop{\ell'=\ell\pm 1}\limits_{\ell'\geq 0}
} 
Y_{n\ell}^{\ell'}(x;\rho_n) \left[ X_{n\ell}^{\ell'}(x;\rho_n) \right]^2
,
\label{BetheLogIntegForm}
\eea
where the summation is taken over $\ell'=\ell \pm 1$ if
$\ell \ge 1$ and 
$\ell'=1$ if $\ell= 0$.
Here, 
\bea
\rho_n = \frac{C_A-2C_F}{2C_F}n
\eea
and
\bea
&&
Y_{n\ell}^{\ell'}(x;\rho_n)=\frac{2304 \max (\ell,\ell')}{n^3(2 \ell+1) }
\frac{(n+\ell)!}{(n-\ell-1)!}
\left\{ \frac{1}{(2 \ell+1)! (2 \ell'+1)!} \right\}^2
\nonumber\\&&~~~~~~~
\times
\frac{x^{2 \ell+1} \rho _n^{2 \ell'+3}}
{\left(\rho
   _n^2+x^2\right)^{\ell+\ell'+2}
}\,
\frac{\exp [4 x \arctan
   \left({\rho _n}/{x}\right)] }{e^{2 \pi  x}-1}
 \Delta _{\ell'}(x) 
 \log \left(\frac{n^2 x^2}{\rho_n^2+x^2}\right)
 ,
\eea
\bea
\Delta_{\ell'}(x) = \prod_{s=1}^{\ell'}(x^2+s^2)
\, .
\eea

Each $X_{n\ell}^{\ell'}$ is a rational function of $x$.
We list explicit forms for some of the lower states:
\begin{eqnarray}
X_{10}^{1} (x;\rho)&=&2+\rho
, \\
X_{20}^{1} (x;\rho)&=&
-
\frac{x^2(8+9\rho +2\rho^2)-\rho^2(4+\rho) }{(\rho^2+x^2)}
, \\
X_{30}^{1} (x;\rho)&=&
\frac{1}{3(\rho^2+x^2)^2}
\bigg\{
x^4 (66+123\rho+60\rho^2+8\rho^3)
\nonumber\\
&&
-2x^2 \rho^2( 54+41\rho+6\rho^2)
+3\rho^4 (6+\rho) 
\bigg\}       
, \\
X_{21}^{2}  (x;\rho)&=&
80(3+\rho)
, \\
X_{21}^{0}  (x;\rho)&=&
\frac{2x^2 (3+9\rho+6\rho^2+\rho^3) -2\rho^2(3+2\rho)}{3(\rho^2+x^2)}
, \\
X_{31}^{2}  (x;\rho)&=&
-
\frac{40x^2(15+12\rho+2\rho^2)-40\rho^2 (9+2\rho)}{(\rho^2+x^2)}
, \\
X_{31}^{0}  (x;\rho)&=&
-
\frac{1 }{3(\rho^2+x^2)^2}
\bigg\{
x^4 (9+42\rho+48\rho^2+18\rho^3+2\rho^4)
\nonumber\\
&&         -2x^2 \rho^2 (15+27\rho+11\rho^2+\rho^3)
         +\rho^4(9+4\rho)
\bigg\}       .
\end{eqnarray} 
Thus, for each $(n,\ell)$,
the integrand of Eq.~(\ref{BetheLogIntegForm}) is given by
a combination of elementary functions, and the integral can be
evaluated easily numerically.
For instance,
\bea
&&
L_{\rm Bethe}(1,0)=\int_0^\infty dx\,
\frac{2312 x \left(x^2+1\right) \log
   \left(\frac{x^2}{x^2+\frac{1}{64}}\right) e^{4 x \arctan
   \left(\frac{1}{8 x}\right)}}{\left(e^{2 \pi  x}-1\right) \left(64
   x^2+1\right)^3}
   \nonumber\\
   &&~~~~~~~~~~~~~~
   =-81.5379 \cdots .
\eea
Numerical values of the Bethe logarithm are given
in Tab.~\ref{TableNumBetheLog}
for some of the lower states.

\begin{table}[htb]
\begin{center}
\begin{tabular}{|l|l|l|l|l|l|} \hline
& ~~$s(\ell=0)$ & ~~$p(\ell=1)$ & ~~$d(\ell=2)$ & ~~$f(\ell=3)$ & ~~$g(\ell=4)$\\ \hline
$n=1$   & $-81.5379$ &&&& \\ 
$n=2$   & $ -37.6710$ &$ -0.754367$ &&& \\ 
$n=3$   & $ -22.4818$ & $+2.04826$ & $+4.62186$ && \\ 
$n=4$  & $ -14.5326$ & $+3.85121$ & $+5.80596$ & $+6.47937$ & \\ 
$n=5$   & $ -9.52642$ & $+5.17957$ & $+6.76019$ & $+7.32254$ & $+7.64759$  \\ 
\hline
\end{tabular}
\end{center}
\caption{\label{TableNumBetheLog}
Numerical values of the Bethe logarithm $L_{\rm Bethe}(n,\ell)$
for some of the lower states.
The color factors are set as $C_F=4/3$ and $C_A=3$.
}
\end{table}

\section{Full Formula}
Combining all the corrections we present the full formula
of the heavy quarkonium energy levels up to NNNLO,\footnote{
A {\it Mathematica} file of the full formula can be
downloaded at \texttt{http://www.tuhep.phys.tohoku.
ac.jp/$\sim$program/NNNLOFullFormula/FullNNNLOFormula-Paper.nb.}
} that is,
up to ${\cal O}(\alpha_s^5 m)$
and ${\cal O}(\alpha_s^5 m \log \alpha_s)$.
The heavy quarkonium state is identified by the
quantum numbers $n$, $\ell$, $s$ and $j$.
In this section we write the pole mass of $Q$ or $\bar{Q}$
as $M_{Q,{\rm pole}}$ instead of $m$, to avoid confusions,
since $m$ is used to represent the summation index corresponding
to $\ell_z$.

The energy level is given by
\bea
&&
M_{Q\bar{Q}}(n,\ell,s,j)=M_{Q,{\rm pole}}
\left[ 2 -\frac{C_F^2\alpha_s^2}{4n^2}\sum_{i=0}^3
\biggl(\frac{\alpha_s}{\pi}\biggr)^i \, P_i(L_\mu)
\right]
, \\ &&
L_\mu = \log\left(\frac{n\mu}{C_F\alpha_s M_{Q,{\rm pole}}}\right)+S_1(n+\ell)
,
\eea
where the coefficients of logarithms of $\mu$ in $P_i$ are 
determined by the renormalization group 
as
\bea
&&
P_0=1
, \\ &&
P_1=\beta_0 L_\mu + c_1
, \\ &&
P_2=\frac{3}{4} \beta _0^2 L_{\mu }^2
+
\left(-\frac{\beta _0^2}{2}+\frac{\beta _1}{4}+\frac{3 \beta _0 c_1}{2}\right)
   L_{\mu }+c_2
, \\ &&
P_3=\frac{1}{2} \beta _0^3 L_{\mu }^3+
\left(-\frac{7 \beta _0^3}{8}+\frac{7 \beta _0 \beta _1}{16}+\frac{3}{2} \beta _0^2
   c_1\right) L_{\mu }^2
\nonumber\\&&~~~~~~~
+\left(\frac{\beta _0^3}{4}-\frac{\beta_0 \beta _1 }{4}+\frac{\beta _2}{16}-\frac{3}{4} \beta _0^2 c_1+2 \beta _0 c_2+\frac{3
   \beta _1 c_1}{8}\right) L_{\mu } +c_3
   .
\eea
The color factors, $C_A,C_F,T_F$,
the coefficients
of the beta function, $\beta_0, \beta_1, \beta_2$, as well as the constants of
the static QCD potential, 
$a_1, a_2, \bar{a}_3$, to be used below
are given in App.~\ref{AppCFandParams}.

Hereafter, we use $\delta_{\ell 0}$ and $\delta_{\ell \ge 1}$(equal to one if
$\ell \ge 1$ and zero if $\ell=0$)
to separate the cases $\ell=0$ and $\ell \ge 1$.
We separate the constant part as $c_i=c_i^{\rm c}+c_i^{\rm nc}$, 
where $c_i^{\rm c}$ corresponds to the
contributions purely from the (renormalized)
Coulomb potential, while 
$c_i^{\rm nc}$ corresponds to the
sum of all the rest of the contributions:
\bea
&&
c_1=\frac{a_1}{2}
, \\ &&
c_2=c_2^{\rm c}+c_2^{\rm nc}
, \\ &&
c_{2}^{\rm c}=\frac{a_1^2}{16}+\frac{a_2}{8}-\frac{\beta _0 a_1 }{4}+\beta _0^2 \, \nu (n,\ell)
, \\ &&
c_2^{\rm nc}=
\pi ^2 C_F^2 \left\{
\frac{2}{ n(2 \ell+1)}
-\frac{11}{16 n^2} 
-\frac{D_S+3 X_{\text{LS}}}{2 n \ell (\ell+1) (2
   \ell+1)}\,\delta
   _{\ell\geq 1} 
-\frac{2  }{3 n}\,\mathbb{S}^2\delta
   _{\ell 0}\right\}
+ \frac{\pi^2 C_F C_A}{ n(2 \ell+1)}
,
\nonumber\\&&~~~~~~~
\\ &&
c_3=c_3^{\rm c}+c_3^{\rm nc}
, \\ &&
c_3^{\rm c}=
\frac{\beta _0^2 a_1 }{8} +\frac{3 \beta_0 a_1^2}{32} 
   -\frac{\beta _0 a_2 }{16}-\frac{ \beta _1 a_1}{16}
-\frac{a_1^3}{16}-\frac{3 a_1 a_2}{32}
   + \frac{\bar{a}_3}{32}
   +a_1
   c_2^{\rm c}
\nonumber\\&&~~~~~~~
+\beta _0 \beta _1\, \sigma (n,\ell)+\beta _0^3 \, \tau
   (n,\ell)+\frac{\pi ^2}{2}  C_A^3 L_{\mu }
, \\ &&
c_3^{\rm nc}=
\pi ^2 \biggl(C_A^3 \,\xi _{\text{AAA}}+C_A^2  C_F\,\xi _{\text{AAF}}+C_A  C_F^2\,\xi
   _{\text{AFF}}+C_F^3 \,\xi
   _{\text{FFF}}
\nonumber\\&&~~~~~~~
~
+C_A  C_F T_Fn_{\ell} \,\xi _{\text{AFnl}}+C_F^2  T_Fn_{\ell} \,\xi _{\text{FFnl}}+C_F^2 T_F \,\xi _{\text{FF}}
-\frac{1}{6}
   \beta _0 n c_2^{\text{nc}}
\biggr)-\frac{\pi ^2}{2}  C_A^3 L_{\mu }
,
\eea
\bea
&&
\nu (n,\ell)=\frac{n}{2} \zeta (3) 
+\frac{\pi ^2 }{8} \left(1-\frac{2 n }{3}
\text{$\Delta
   $}S_{\rm 1a}
\right)-\frac{1}{2} S_2(n+\ell)
+\frac{n}{2}\Sigma_{\rm a}(n,\ell)
, \\ &&
\sigma (n,\ell)=\frac{\pi ^2}{64}-\frac{1}{16} S_2(n+\ell)
+\frac{1}{8}\Sigma^{(k)}_2+\frac{1}{2}\nu (n,\ell)
   , \\ &&
\tau (n,\ell)=
\frac{3}{2} \zeta (5) n^2-\frac{ \pi ^2}{8} \zeta (3) n^2
+
\frac{\pi ^4  }{1440}n\left(5 n \text{$\Delta $}S_{\rm 1a}-4\right)
\nonumber\\&&~~~~~~~
-\frac{1}{4} \zeta (3) \left[\left(n \text{$\Delta $}S_{\text{1a}}-2\right)^2+n^2
   \left\{2 S_2(n+\ell)-S_2(n-\ell-1)\right\}+n-4\right]
\nonumber\\&&~~~~~~~
+\frac{\pi ^2}{12}  \biggl[\frac{n}{2}  \text{$\Delta $}S_{\text{1a}} \left\{n
   S_2(n+\ell)+1\right\}+\frac{n^2}{2}  S_3(n+\ell)-\frac{3}{4}
   -n^2 \Sigma_{\rm a}(n,\ell) \biggr]
\nonumber\\&&~~~~~~~
-\frac{n^2}{2}  S_{4,1}(n-\ell-1)+n S_{3,1}(n-\ell-1)+\frac{1}{4} S_2(n+\ell)+\frac{1}{2}
   S_3(n+\ell)
\nonumber\\&&~~~~~~~
+\Sigma_{\tau,1}(n,\ell)+\Sigma_{\tau,2}(n,\ell)+\Sigma_{\tau,3}(n,\ell)
.
\eea
Definitions of various finite sums are given as follows.
\bea
&&
S_p(N)=\sum_{i=1}^N\frac{1}{i^p},
~~~~~
S_{p,q}(N)=\sum_{i=1}^N\sum_{j=1}^i\frac{1}{i^pj^q},
\\ &&
\Delta S_{\rm 1a}=S_1(n+\ell)-S_1(n-\ell-1) 
, \\ &&
\Delta S_{\rm 1b}=S_1(n+\ell)-S_1(2\ell+1)
,
\eea
\bea
&&
\Sigma_{\rm a}(n,\ell)=
\Sigma^{(m)}_3 +\Sigma^{(k)}_3+\frac{2}{n}\Sigma^{(k)}_2
, \\ &&
\Sigma_{\rm b}(n,\ell)=\Sigma^{(m)}_2+\Sigma^{(k)}_2-\frac{2}{n}\Delta S_{\rm 1b}
,
\eea
\bea
&&
\Sigma^{(m)}_{p}(n,\ell)=
\frac{(n+\ell)!}{(n-\ell-1)!} \sum _{m=-\ell}^\ell \frac{R(\ell,m)}{(n+m)^p}S_1(n+m) 
, \\ &&
\Sigma^{(k)}_{p}(n,\ell)=
\frac{(n-\ell-1)!   }{(n+\ell)!}
\sum _{k=1}^{n-\ell-1} \frac{(k+2 \ell)!}{(k-1)!
   (k+\ell-n)^p}
   ,
\eea
\bea
R(\ell,m)=\frac{(-1)^{\ell-m}}{(\ell+m)!(\ell-m)!}
,
\eea
\bea
&&
\Sigma_{\tau,1}=
-\frac{n^2 (n+\ell)!}{4 (n-\ell-1)!}
 \sum _{k=1}^{n-\ell-1} \frac{(k-1)!
   S_1(n-\ell-k)}{(k+2 \ell)! (k+\ell-n)^4}
\nonumber\\&&~~~~~~~
   +
\frac{(n-\ell-1)! }{4
   (n+\ell)!}
   \sum _{k=1}^{n-\ell-1} \frac{(k+2 \ell)! }{(k-1)! (k+\ell-n)^4}   
\nonumber\\&&~~~~~~~
\times
      \Biggl[ (k+\ell-n) (2 k+2 \ell-n)
   \left\{2 n S_2(n-\ell-k-1)-1\right\}
\nonumber\\&&~~~~~~~~~~
-6
   \left\{(k+\ell-n) (2 k+2 \ell-n)+n \left(k+\ell-\frac{n}{3}\right)\right\} S_1(n-\ell-k-1)
\nonumber\\&&~~~~~~~
~~~
+\{3
   (k+\ell-n) (2 k+2 \ell-n)+n (k+\ell)\} \left\{S_1(k+2 \ell)-S_1(n+\ell)\right\}
\Biggr]
, \\ &&
\Sigma_{\tau,2}=
\frac{n (n+\ell)! }{8 (n-\ell-1)!}
\sum _{m=-\ell}^\ell \frac{R(\ell,m)}{(n+m)^5}
\nonumber\\&&~~~~~~~
\times
\Bigl[ 
4 n (n+m)^2 S_{2,1}(n+m)
-(4 m+3 n) (n+m)
   S_2(n+m)
\nonumber\\&&~~~~~~~
~~~
+S_1(n+m) \bigl\{ 
-2 (n+m)^2-8 n +
8 (n+m)^2 S_1(2 \ell+1)-2 n (n+m) S_1(\ell+m)
\nonumber\\&&~~~~~~~
~~~
-2 (4
   m+3 n) (n+m) S_1(\ell+n)-(4 m-n) (n+m) S_1(n+m)\bigr\}
\Bigr]
, \\ &&
\Sigma_{\tau,3}=
n^2 \sum _{m=-\ell}^{\ell} \sum _{k=1}^{n-\ell-1} \frac{(k+2 \ell)!
   S_1(n+m) R(\ell,m) }{(k-1)!
   (n+m)^2 (k+\ell+m)}
   \left\{\frac{1}{2 (k+\ell-n)^2}-\frac{1}{n (n+m)}\right\}
   .
\eea
Non-Coulomb corrections are classified according to different
color factors:
\bea
&&
\xi _{\text{AAA}}=
-\frac{5}{36}+
\frac{1}{6}L_{\text{US}}
, \\ &&
\xi _{\text{AAF}}=
\frac{5 (2 \ell-3)}{4 n (2 \ell+1)^2}
+\frac{4
   }{3 n (2 \ell+1)}L_{\text{US}}+
\frac{1}{3 n (2 \ell+1)}(11 n \Sigma_{\rm b}-8 \text{$\Delta $}S_{\text{1b}})
, \\ &&
\xi _{\text{FFF}}=
-\frac{2}{3} L_{\text{Bethe}}
-\frac{ D_S+2
   X_{\text{LS}}-{7}/{3}}{2n \ell (\ell+1) (2 \ell+1)
   }\,\delta _{\ell\geq 1}
\nonumber\\&&~~~~~~~~~
+\frac{\delta _{\ell 0}}{3 n}
 \left\{8 L_{\text{US}}-L_{\rm H}-14 S_1(n)+\frac{7}{2
   n}+\mathbb{S}^2-\frac{79}{6}\right\}
,
\\ &&
\xi _{\text{FF}}=
\frac{\delta _{\ell 0} }{n}
\left\{\frac{32}{15}-\mathbb{S}^2+(\mathbb{S}^2-2) \log
   2\right\}
, \\ &&
\xi _{\text{FFnl}}=
\frac{11 \pi ^2}{72 n}+\frac{37}{36 n^2}-\frac{62
   \ell+7}{9 n (2 \ell+1)^2}-\frac{8 \Sigma_{\rm b}}{3 (2 \ell+1)}
\nonumber\\&&~~~
~~~~~
   +\frac{2 \mathbb{S}^2
  }{27
   n^2} \delta _{\ell 0} \left\{12 n^2 S_2(n)-24 n S_1(n)+11 n+3\right\}
+
\frac{\delta _{\ell\geq 1}}{ n\ell (\ell+1) (2 \ell+1)}
\Biggl[ \,
 \frac{2}{9}
   \mathbb{S}^2
\nonumber\\&&~~~~~~~~
~~~
-\frac{D_S+2 X_{\text{LS}}}{3}
+
\left(D_S+3 X_{\text{LS}}\right)
   \left\{\frac{2 \ell+1}{6 n}
-\frac{4 \ell^2+6 \ell+1}{6 \ell (\ell+1) (2
   \ell+1)}+\frac{2 n \Sigma_{\rm b}}{3}\right\}
\Biggr]
,
\\ &&
\xi _{\text{AFnl}}=
\frac{1}{2
   \ell+1}
   \left\{
   -\frac{130 \ell+17}{36 n (2 \ell+1)}-\frac{4 \Sigma_{\rm b}}{3}
   \right\}
, \\ &&
\xi _{\text{AFF}}=
-\frac{7}{16 n^2}-\frac{121 \pi ^2}{288 n}+
\frac{22 n \Sigma_{\rm b}-32 \text{$\Delta $}S_{\text{1b}}}{3 n (2 \ell+1)}
+
L_{\text{US}} \left\{\frac{16}{3 n (2 \ell+1)}-\frac{4 }{3
   n}\delta _{\ell 0}-\frac{2}{3 n^2}\right\}
\nonumber\\&&~~~~~~~
   +
   L_{\rm H} \left\{\frac{ D_S+2 X_{\text{LS}}}{2 n
   \ell (\ell+1) (2 \ell+1)}\delta _{\ell\geq 1}
+\frac{7 \mathbb{S}^2-17}{6 n} \delta _{\ell 0}\right\}
+\frac{\delta _{\ell\geq 1}}{ n\ell (\ell+1) (2 \ell+1)}
\nonumber\\&&~~~~~~~
   ~~\times\!
 \Biggl[
 -\frac{94 \ell^3+597
   \ell^2+653 \ell+75}{36 (2 \ell+1)}
    -\frac{\mathbb{S}^2}{36}
   +
   \left(D_S+2 X_{\text{LS}}\right)
 \left\{\frac{1}{2 (2 \ell+1)}-\text{$\Delta $}S_{\text{1b}}-\frac{5}{33}\right\}
\nonumber\\&&~~~~~~~
~~~~~
   +
 \left(5 D_S+21 X_{\text{LS}}\right)
\left\{\frac{4 \ell^2+6 \ell+1}{24
   \ell (\ell+1) (2 \ell+1)}-\frac{2 \ell+1}{24
   n}+\frac{13}{66}\right\}
 -\frac{11}{6} n \Sigma_{\rm b} \left(D_S+3 X_{\text{LS}}\right)
   \Biggr]
\nonumber\\&&~~~~~~~
+
\frac{\delta _{\ell 0}}{3 n}
 \left\{-\frac{22}{3} n \mathbb{S}^2 S_2(n)+\left(\frac{23
   \mathbb{S}^2}{3}+25\right) S_1(n)+\left(-\frac{1}{12 n}-\frac{95}{36}\right)
   \mathbb{S}^2-\frac{25}{4 n}-\frac{106}{3}\right\}
   ,
\nonumber\\&&~~~~~~~
\eea    
\bea
&&
L_{\rm H} = \log\left(\frac{n}{C_F\alpha_s}\right)+S_1(n+\ell)
, \\ &&
L_{\rm US} = \log\left(\frac{nC_F\alpha_s}{2}\right)+S_1(n+\ell)
,
\eea

\bea
&&
\mathbb{S}^2\equiv \left< \vec{S}^2 \right> =s(s+1)
,
\\ &&
D_{S} \equiv
\left< 
3 \frac{(\vec{r}\cdot \vec{S})^2}{r^2} - \vec{S}^2 
\right>
=
\frac{
2 \ell (\ell+1) s (s+1) - 3 X_{LS} - 6 X_{LS}^2
}{
(2\ell-1)(2\ell+3)
},
\\&&
X_{LS} \equiv
\left< \vec{L}\cdot \vec{S} \right>
= \frac{1}{2}\,
\left[ j(j+1)-\ell(\ell+1)-s(s+1) \right] .
\eea

The Bethe logarithm is given as a one-parameter integral:
\bea
L_{\rm Bethe}(n,\ell)=\int_0^\infty dx
\sum_{
\mathop{\ell'=\ell\pm 1}\limits_{\ell'\geq 0}
} 
Y_{n\ell}^{\ell'}(x;\rho_n) \left[ X_{n\ell}^{\ell'}(x;\rho_n) \right]^2
,
\eea
where $X_{n\ell}^{\ell'}$ and $Y_{n\ell}^{\ell'}$ are defined as 
\bea
&&
384 \left(\frac{x^2}{\rho_n
   ^2+x^2}\right)^{\frac{1}{2} (\ell+\ell'+5)}
   \exp\left[{2 x \arctan\left(\frac{\rho_n }{x}\right)}\right] 
   X_{n\ell}^{\ell'}(x;\rho_n)
\nonumber\\&&~~~
   =
(2 \ell+1)! (n-\ell-1)! 
\sum _{k=0}^{n-\ell-1} \frac{(-1)^k
    (k+\ell+\ell'+3)! 
   }{k! (k+2 \ell+1)! (n-k-\ell-1)!}\,
   \left(\frac{2x}{x+i  \rho_n
   }\right)^{k+\ell+\ell'+4} \,
\nonumber\\&&~~~~~~~
\times
   {_2F_1}\left(\ell'+1-{i}{x},k+\ell+\ell'+4;2 \ell'+2;\frac{2 i \rho_n
   }{x+i  \rho_n}\right)
   ,
\eea
\bea
&&
Y_{n\ell}^{\ell'}(x;\rho_n)=\frac{2304 \max (\ell,\ell')}{n^3(2 \ell+1) }
\frac{(n+\ell)!}{(n-\ell-1)!}
\left\{ \frac{1}{(2 \ell+1)! (2 \ell'+1)!} \right\}^2
\nonumber\\&&~~~~~~~
\times
\frac{x^{2 \ell+1} \rho _n^{2 \ell'+3}}
{\left(\rho
   _n^2+x^2\right)^{\ell+\ell'+2}
}\,
\frac{\exp [4 x \arctan
   \left({\rho _n}/{x}\right)] }{e^{2 \pi  x}-1}
 \Delta _{\ell'}(x) 
 \log \left(\frac{n^2 x^2}{\rho_n^2+x^2}\right)
 ,
\eea
with 
\bea
\rho_n = \frac{C_A-2C_F}{2C_F}n ,
~~~~~~
\Delta_{\ell'}(x) = \prod_{m=1}^{\ell'}(x^2+m^2)
\, .
\eea
The hypergeometric function is defined as
$
_2F_1(a,b;c;z)=\frac{\Gamma(c)}{\Gamma(a)\Gamma(b)}
\sum_{n=0}^\infty \frac{\Gamma(a+n)\Gamma(b+n)}{\Gamma(c+n)}
\frac{z^n}{n!}
$.
Each
$ X_{n\ell}^{\ell'}(x;\rho_n)$ is a rational function of $x$.
Explicit expressions of $X_{n\ell}^{\ell'}(x;\rho_n)$
and
numerical values of the Bethe logarithm
$L_{\rm Bethe}(n,\ell)$ are given
in Sec.~\ref{sec:BetheLog},
for some of the lower states.

The above formula reproduces the known results:
(1) the coefficients of
$\alpha_s^5 m \log \alpha_s$ for general quantum numbers $(n,\ell,s,j)$
\cite{Brambilla:1999xj}, and (2)
the NNNLO formula for the general $S$-wave state  
$(n,j)$ \cite{Beneke:2005hg}.

\section{Summary and Discussion}

We have computed the energy levels of the heavy quarkonium states
in a double expansion in $\alpha_s$ and $\log\alpha_s$
up to order $\alpha_s^5 m$ and $\alpha_s^5 m \log\alpha_s$.
The result is given as a general formula dependent on the
quantum numbers $(n,\ell,s,j)$ of the bound state.
The computation is performed using the theoretical framework
pNRQCD.

In pNRQCD
the perturbative corrections consist of contributions from
the potential and 
ultra-soft regions.
Decomposition of both contributions
is realized by a multipole expansion.
The corrections from the potential region
(potential corrections)
are identical to perturbative corrections of
energy levels
in quantum mechanics.
The relevant Hamiltonian has been known,
and it is straightforward to obtain an infinite sum formula
using a known infinite-sum representation of the Green function.
We used a recent technology for evaluating
multiple sums to reduce infinite
sums to combinations of transcendental numbers and finite sums.

Corrections involving the ultra-soft region (ultra-soft corrections)
start from the order $\alpha_s^5 m$ and $\alpha_s^5 m \log\alpha_s$.
The relevant correction in our computation is given by
a one-loop
self-energy of a singlet $Q\bar{Q}$ state by emitting
and reabsorbing an ultra-soft gluon.
It can be separated to a part given by
an expectation value of a Hamiltonian (local in time)
and a part
given by a contribution non-local in time
(QCD Bethe logarithm).
A UV divergence is included in the Hamiltonian,
which cancels the IR divergence contained in the
potential corrections.
A new feature of our computation is that
these are computed in an arithmetic manner, in contrast to
previous computations which
involve diagrammatic analysis.
Hence, our method may be easier to follow for
non-experts.
The Bethe logarithm is given as a one-parameter integral.
Its numerical values for some lower states are
given.
All the other parts are evaluated analytically, besides
$\bar{a}_3$ which is as yet known only numerically.

The obtained formula is quite lengthy.
One reason is that part of the formula needs to be
given separately 
for the cases $\ell=0$ and $\ell>0$.
Another reason is that there appear various different
types of finite sums.
(They may be reduced to more compact expressions if we understand the nature
of the finite sums better.)

In general,
there are two ways to compute the energy levels
using perturbative QCD.
One way is to compute thoroughly within perturbative QCD.
(Technically this is done efficiently using an EFT.)
The other way is to compute by factorizing UV and IR contributions in 
an operator-product expansion.
In the former computation, there are well-established prescriptions to estimate
uncertainties of the prediction within perturbative QCD.
Empirically estimates of 
perturbative uncertainties are approximated well by IR renormalons
of order $\LQ^3a_X^2$, where $a_X$ denotes the typical radius of
the bound state $X$.
In the latter computation, UV contributions are encoded in the Wilson 
coefficients, which are free from IR renormalons, while IR contributions
are included in non-perturbative matrix elements.
The correspondence of the two computations is that IR part of the former 
computation is replaced by the matrix elements of the latter, and that the 
residual UV contributions of the former equals the Wilson coefficients of
the latter.
Thus, the uncertainties by IR renormalons $\sim \LQ^3a_X^2$
in the former computation are
replaced by the non-perturbative matrix elements in the latter computation.
Our computation in this paper corresponds to the former type of computation. 
See discussion in \cite{Kiyo:2013aea} for more details in the case of 
the bottomonium spectrum.

\section*{Acknowledgments}

The authors are grateful to S.~Recksiegel for pointing out 
a misprint in App.~A.
The works of Y.K.\ and Y.S., respectively,
were supported in part by Grant-in-Aid for
scientific research Nos.\ 26400255 and 26400238 from
MEXT, Japan.


\vspace*{10mm}
\appendix
\clfn
\section*{Appendices}

\section{Conventions and Notations}
\label{AppCFandParams}

In this appendix we present definitions of parameters
and conventions used in the main body of this paper.

The color factors for the $SU(N_C)$ gauge group are given by
\bea
&&
T_F=\frac{1}{2},
~~~
C_F=\frac{N_C^2-1}{2N_C},
~~~
C_A=N_C,
\\ &&
\frac{d_F^{abcd}d_F^{abcd}}{N_A\,T_F}=\frac{N_C^4-6N_C^2+18}{48 N_C^2},
~~~
\frac{d_A^{abcd}d_F^{abcd}}{N_A\,T_F}=\frac{N_C(N_C^2+6)}{24}
.
\eea
$N_C=3$ for QCD.

The coefficients of the beta function are given by
\bea
&&
\beta_0=\frac{11}{3}\,C_A-\frac{4}{3}\,n_l\,T_F\,,
~~~
\beta_1=\frac{34}{3}\,C_A^{\,2}
-\left(\frac{20}{3}\,C_A+ 4\,C_F \right)\,n_l T_F\,,
\nonumber\\
&&
\beta_2=
\frac{2857}{54}\,C_A^{\,3}
-\left( \frac{1415}{27}\,C_A^2
       +\frac{205}{9}\, C_A C_F
       -2\,C_F^2
\right)\, n_l T_F
\nonumber\\
&& ~~~~~~~
+\left(\frac{158}{27}\,C_A+\frac{44}{9}\,C_F\right)\, n_l^2 T_F^{\,2} 
\, ,
\eea
where $n_l$ denotes the number of massless quark flavors.

The constants in the static QCD potential are given by
\begin{eqnarray}
&&
a_0=1\,,
~~~
a_1 =
\frac{31}{9}\,C_A -\frac{20}{9}\,T_F\,n_l\,,
\nonumber\\&&
a_2 =
 \left(\frac{4343}{162}+4\pi^2-\frac{\pi^4}{4}+\frac{22}{3}\zeta_3\right)\,C_A^{\,2}
-\left( \frac{1798}{81}+\frac{56}{3}\zeta_3\right)\,C_A T_F n_l
\nonumber \\ &&
~~~~~~
-\left(\frac{55}{3}-16\zeta_3 \right)\,C_F T_F n_l
+\left(\frac{20}{9}T_F n_l\right)^2\,,
\label{eq:a1-a2}
\\
\bar{a}_3
&=&
-\left(\frac{20}{9}n_l T_F\right)^3 \!
+\bigg[
       C_A\left( \frac{12541}{243}
                +\frac{64\pi^4}{135}
                +\frac{368}{3}\zeta_3\right)
   +C_F\left( \frac{14002}{81}
                -\frac{416}{3}\zeta_3\right)
 \bigg]\,(n_l T_F)^2
\nonumber \\
&+&
\bigg[\,
  2\,\gamma_1\, C_A^{\,2}
 +\left(-\frac{71281}{162}+264\zeta_3+80\zeta_5\right)C_A\, C_F
 +
\left(\frac{286}{9}+\frac{296}{3}\zeta_3-160\zeta_5\right) C_F^{\,2}
\bigg]\,n_l\,T_F
\nonumber \\
&+&
\frac{1}{2}\,\gamma_2\,n_l\, \biggl(\frac{d_F^{abcd}d_F^{abcd}}{N_A\,T_F}\biggr)\,
+\bigg[\,
\gamma_3\, C_A^{\,3}
+\frac{1}{2}\,\gamma_4\, \biggl(\frac{d_A^{abcd}d_F^{abcd}}{N_A\,T_F}\,\biggr)
\bigg]
\,,
\end{eqnarray}
In the above equation, the coefficients
$\gamma_i$'s are known only numerically:
\bea
&&
\gamma_1 = -354.859, ~~~ \gamma_2=-56.83(1), ~~\cite{Smirnov:2008pn}
\eea
and\footnote{
The results of Ref.~\cite{Smirnov:2009fh} are
$\gamma_3=502.24(1)$, $\gamma_4=-136.39(12)$.
}
\bea
&&
\gamma_3=502.22(12) 
\,,
~~~
\gamma_4=-136.8(14) ~~\cite{Anzai:2009tm}
\,.
\eea

The strong coupling
constant in the $\overline{\rm MS}$ scheme 
is defined in the following way.
The bare gauge coupling constant is expressed as
\bea
g_0=\biggl(\frac{\mu^2 e^{\gamma_E}}{{4\pi}}\biggr)^{\epsilon/2} Z_g g_R
=\bar{\mu}^\epsilon Z_g g_R.
\eea
We choose the counter terms to be 
only multiple poles in $\epsilon$ in dimensional
regularization:
\bea
Z_g=1+\sum_{n=1}^\infty \frac{z_n(g_R)}{\epsilon^n} ,
\eea 
where each $z_n(g_R)$ is given as a series expansion in $g_R$.
Then the strong coupling constant in the $\overline{\rm MS}$ scheme 
is given by 
\bea
\alpha_s(\mu)\equiv \frac{g_R^2}{4\pi} .
\eea

\section{Formulas for Fourier Transformation}
\label{app1}

We can use the following formulas for Fourier transforms
in $d=3-2\epsilon$ spatial dimensions:
\bea
&&
F(r,\mu;u)= \int \frac{d^{d}\vec{q}}{(2\pi)^{d}}\, 
\frac{e^{i\vec{q}\cdot\vec{r}}}{q^2}\left(\frac{\mu^2}{q^2}\right)^u
=
\frac{4^{-u-1} \pi ^{\epsilon -\frac{3}{2}} \Gamma \left(-u-\epsilon +\frac{1}{2}\right)}{\Gamma (u+1)}\,\mu ^{2 u} r^{2 u+2 \epsilon
   -1} ,
   \label{FTformula1}
   \\&&
G(q,\mu;u)= \int {d^{d}\vec{r}}\, \,
\frac{e^{-i\vec{q}\cdot\vec{r}}}{r}\left({\mu^2}{r^2}\right)^u
=
\frac{4^{u-\epsilon +1}
\pi ^{\frac{3}{2}-\epsilon } \Gamma (u-\epsilon +1)}{\Gamma \left(\frac{1}{2}-u\right)}
   \,\mu ^{2 u}  q^{-2 u+2   \epsilon -2} ,
\eea
where $q=|\vec{q}|$ and $r=|\vec{r}|$.
If we differentiate $F$ and $G$ by $\vec{r}$ and $\vec{q}$, 
respectively, we may also obtain formulas for
Fourier transforms of tensor operators.

\section{Coulomb Matrix Elements}
\label{app:CoulombME}

\subsection{Formulas}
The generating function of the Laguerre polynomial is
given by
\begin{eqnarray}
U_a(z,t)=\frac{e^{-\frac{t z}{1-t}}}{(1-t)^{1+a}}=\sum_{m=0}^\infty L_m^a(z) t^m
.
\label{GeneFnLaguerre}
\end{eqnarray}
It is easy to evaluate an integral of the generating functions
\begin{eqnarray}
\int_0^\infty dz\, z^{a+\beta} e^{-z} U_a(z,t) U_a(z,s) 
&=&
\Gamma(1+a+\beta)\frac{(1-t)^\beta (1-s)^\beta}{ (1-st)^{1+a+\beta}}
.
\label{eq:integration_generating_function_LaguerreL}
\end{eqnarray}
Expansions in $s$ and $t$ read
\begin{eqnarray}
&&
(1-t)^\beta 
=
\sum_k \frac{(-1)^k \Gamma(1+\beta)}{\Gamma(1+\beta-k) k!} \, t^k
=
\sum_k \frac{\Gamma(k-\beta)}{\Gamma(-\beta) k!} t^k
,
\\ &&
(1-st)^{-1-a-\beta} 
=\sum_k \frac{\Gamma(k+1+a+\beta)}{\Gamma(1+a+\beta) k!} (st)^k
.
\end{eqnarray}

\subsubsection*{Diagonal part}

If we take the 
diagonal part of Eq.~(\ref{eq:integration_generating_function_LaguerreL}),
\begin{eqnarray}
 U_a(z,t) U_a(z,s)\biggr|_{\rm diagonal } 
&=& 
\sum_{m=0}^\infty L_{m}^a(z) L_{m}^a(z) (st)^m ,
\\
\frac{(1-t)^\beta (1-s)^\beta 
     }{ (1-st)^{1+a+\beta}}\biggr|_{\rm diagonal } 
&=&
\sum_{k=0}^\infty
\frac{\Gamma(k+1+a+\beta)}{\Gamma(1+a+\beta) k!} (st)^k
\times
\sum_{\ell=0}^\infty 
\bigg\{ \frac{\Gamma(\ell-\beta)}{\Gamma(-\beta) \ell!}\bigg\}^2 (st)^\ell
\nonumber
\\
&=&
\sum_{m=0}^\infty
(st)^m
\sum_{k=0}^{m}
\frac{\Gamma(k+1+a+\beta)}{\Gamma(1+a+\beta) k!} 
\bigg\{ \frac{\Gamma(m-k-\beta)}{\Gamma(-\beta) (m-k)!}\bigg\}^2 ,
\end{eqnarray}
we obtain
\begin{eqnarray}
\int_0^\infty dz\, z^{a+\beta}e^{-z} \big[ L_m^a(z) \big]^2
&=&
\sum_{k=0}^m
\frac{\Gamma(1+a+\beta+k)}{k!} 
\bigg\{ \frac{\Gamma(m-k-\beta)}{\Gamma(-\beta) (m-k)!}\bigg\}^2.
\end{eqnarray}
By taking the limits $\beta\to -2, -1, 0, \cdots $, the following 
relations are obtained.
\begin{eqnarray}
\int_0^\infty dz\, z^{a-2} e^{-z} \big[ L_m^a(z) \big]^2
&=&
\sum_{k=0}^m \frac{(1+m-k)^2 \Gamma(a+k-1)}{\Gamma(1+k)} 
\nonumber\\
&=&
\frac{a(a+1)+m(1+3a+2m)}{(a+1)a(a-1)}
\frac{\Gamma(a+m)}{\Gamma(1+m)}
\label{DiagLaguerreFormula1}
, \\
\int_0^\infty dz\, z^{a-1} e^{-z} \big[ L_m^a(z) \big]^2
&=&
\sum_{k=0}^m \frac{\Gamma(a+k)}{\Gamma(1+k)} 
=
\frac{\Gamma(1+a+m)}{a\,\Gamma(1+m)}
, \\
\int_0^\infty dz\, z^{a+0}e^{-z} \big[ L_m^a(z) \big]^2
&=& \frac{\Gamma(1+a+m)}{m!} 
, \\
\int_0^\infty dz\, z^{a+1}e^{-z} \big[ L_m^a(z) \big]^2
&=& \frac{(1+a+2m)\Gamma(1+a+m)}{m!} 
, \\
\int_0^\infty dz\, z^{a+2}e^{-z} \big[ L_m^a(z) \big]^2
&=& \frac{\bigg\{ (1+a)(2+a)+6m (1+a+m) \bigg\}\Gamma(1+a+m)}{m!} 
.
\label{DiagLaguerreFormula-1}
\end{eqnarray}

\subsubsection*{Off-diagonal part}
Expansion of Eq.~(\ref{eq:integration_generating_function_LaguerreL})
reads
\begin{eqnarray} 
&&
\hspace{-1cm}
\sum_{n=0}^\infty\sum_{m=0}^\infty  t^n s^m\,
\int_0^\infty dz\, z^{a+\beta} e^{-z} L_n^a(z) L_m^a(z)
\nonumber\\
&=&
\int_0^\infty dz\, z^{a+\beta} e^{-z} U_a(z, t) U_a(z,s)
\nonumber\\
&=&
\Gamma(1+a+\beta)\frac{(1-t)^\beta (1-s)^\beta}{ (1-st)^{1+a+\beta}}
\nonumber\\
&=&
\sum_{k=0}^\infty\frac{\Gamma(1+a+\beta+k)}{k!} (st)^k
\sum_{i=0}^\infty \frac{\Gamma(i-\beta)}{\Gamma(-\beta) i!} t^i
\sum_{j=0}^\infty \frac{\Gamma(j-\beta)}{\Gamma(-\beta) j!} s^j.
\end{eqnarray}
The off-diagonal $(t^n s^m)$ component corresponds to $i+k=n, j+k=m$, 
with constraints $k \leq n \land k\leq m \land n \neq m$. This gives
\begin{eqnarray}
\int_0^\infty dz\, z^{a+\beta} e^{-z} L_n^a(z) L_m^a(z)
&=&
\sum_{k=0}^{\min(n,m)} \frac{\Gamma(1+a+\beta+k)}{k!} 
 \frac{\Gamma(n-k-\beta)}{\Gamma(-\beta) (n-k)!} 
 \frac{\Gamma(m-k-\beta)}{\Gamma(-\beta) (m-k)!} 
 .
 \nonumber\\
\end{eqnarray}
We derive some formulas assuming $n>m$.
\begin{eqnarray}
&&\hspace{-1cm}
\int_0^\infty dz\, z^{a+\epsilon} e^{-z}  L_n^a(z) L_m^a(z)\biggr|_{n>m}
\nonumber\\
&=&
\sum_{k=0}^{m} \frac{\Gamma(1+a+\epsilon+k)}{k!} 
 \frac{\Gamma(n-k-\epsilon)}{\Gamma(-\epsilon) (n-k)!} 
 \frac{\Gamma(m-k-\epsilon)}{\Gamma(-\epsilon) (m-k)!} 
 \nonumber\\
 &=&
 \sum_{k=0}^{m} \frac{\Gamma(1+a+\epsilon+k)}{k!} 
\bigg\{-\frac{\epsilon}{n-k} +\frac{\epsilon^2(\psi(n-k)+\gamma_E)}{n-k}
+{\cal O}(\epsilon^3)
\bigg\}
\nonumber\\
&&\times
 \bigg \{ 
 \delta_{k,m}-\frac{\epsilon\, \delta_{m\ge k+1} }{m-k}+{\cal O}(\epsilon^2) 
 \bigg\}
 \nonumber\\
 &=&
\epsilon 
\bigg\{-\frac{(a+m)!}{m! (n-m)}
\bigg\}
\nonumber\\
&&
+\epsilon^2
 \bigg\{\frac{(a+m)!\left(\psi(n-m)+\gamma_E -\psi(1+a+m)\right) }{m! (n-m)} 
 +\sum_{k=0}^{m-1} \frac{(a+k)!}{k! (n-k)(m-k)}
 \bigg\}
 \nonumber\\&&
 +{\cal O}(\epsilon^2),
\label{OffdiagLaguerreFormula1}
\end{eqnarray}

\begin{eqnarray}
&&\hspace{-1cm}
\int_0^\infty dz\, z^{-1+a+\epsilon} e^{-z}  L_n^a(z) L_m^a(z)\biggr|_{n>m}
\nonumber\\
&=&
\sum_{k=0}^{m} \frac{\Gamma(a+k+\epsilon)}{k!} 
 \frac{\Gamma(n-k+1-\epsilon)}{\Gamma(1-\epsilon) (n-k)!} 
 \frac{\Gamma(m-k+1-\epsilon)}{\Gamma(1-\epsilon) (m-k)!} 
 \nonumber\\
 &=&
 \frac{(m+a)!}{m! \, a}
 -\epsilon
\sum_{k=0}^m \frac{(k+a-1)!}{k!}
\bigg\{\psi(1+n-k)+\psi(1+m-k)-\psi(a+k)+2\gamma_E\bigg\}
 \nonumber\\
&&
+{\cal O}(\epsilon^2)
,
\end{eqnarray}

\begin{eqnarray}
&&
\int_0^\infty dz\, z^{-2+a+\epsilon} e^{-z}  L_n^a(z) L_m^a(z)\biggr|_{n>m}
\nonumber\\ &&
~~~~~~~
=\sum_{k=0}^{m} \frac{\Gamma(-1+a+k+\epsilon)}{k!} 
 \frac{\Gamma(n-k+2-\epsilon)}{\Gamma(2-\epsilon) (n-k)!} 
 \frac{\Gamma(m-k+2-\epsilon)}{\Gamma(2-\epsilon) (m-k)!} 
 \nonumber\\
 &&~~~~~~~
 =\frac{(m+a)!\left\{1+n+m+a(1+n-m)\right\} }{m! \, a (a+1)(a-1) }
 \nonumber\\
 &&~~~~~~~~~~
 -\epsilon
\sum_{k=0}^m \frac{(k+a-2)! (1+n-k)(1+m-k)}{k!}
\nonumber\\
&&
~~~~~~~~~~~~~
\times
\bigg\{\psi(2+n-k)+\psi(2+m-k)-\psi(a+k-1)+2\gamma_E-2\bigg\}
\nonumber\\
&&~~~~~~~
+{\cal O}(\epsilon^2)
.
\label{OffdiagLaguerreFormula-1}
\end{eqnarray}

As an application we reproduce an orthogonality relation of
the Laguerre polynomial 
\begin{eqnarray}
\int_0^{\infty} dz\,  z^a e^{-z} L_{n}^a(z) L_{m}^a(z) 
&=& \frac{\Gamma(1+a+m)}{m!} \delta_{nm},
\end{eqnarray}
which follows from absence of the off-diagonal part of 
the integration
\begin{eqnarray}
\int_0^\infty dz\, z^{a} e^{-z} U_a(z,t) U_a(z,s) 
&=&
\frac{\Gamma(1+a+\beta)}{ (1-st)^{1+a+\beta}} .
\end{eqnarray}

\subsection{Simple potentials}
Some explicit results for
simple potentials
are listed.
\begin{eqnarray}
\bigg\langle  \frac{1}{r^u} \bigg\rangle_{n\ell}
&=&
\int_0^\infty dr \, r^{2-u}  R_{n\ell}(r)^2
\nonumber\\
&=&
\left(\frac{2}{na_0}\right)^u
\frac{(n-\ell-1)!}{2n (n+\ell)!}
\int_0^\infty dz \, z^{a-u+1} \left(L_m^a(z)\right)^2 
\nonumber\\
&=&
\left(\frac{2}{na_0}\right)^{u}
\frac{(n-\ell-1)!}{2n (n+\ell)!}
\sum_{k=0}^m
\frac{\Gamma(2+a-u+k)}{k!} 
\bigg\{ \frac{\Gamma(m+u-1-k)}{\Gamma(u-1) (m-k)!}\bigg\}^2,
\end{eqnarray}
with $a=2\ell+1, ~~m=n-\ell-1$.
For $u=-1,-2,-3$, it reads:
\begin{eqnarray}
\bigg\langle  \frac{1}{r} \bigg\rangle_{n\ell}
&=&
\frac{1}{2n}
\left(\frac{2}{na_0}\right) ,
\label{eq:vev_pot_r1}
\\
\bigg\langle  \frac{1}{r^2} \bigg\rangle_{n\ell}
&=&
\frac{1}{2n}
\left(\frac{2}{na_0}\right)^2
\frac{1}{(2\ell+1)} ,
\label{eq:vev_pot_r2}
\\
\bigg\langle  \frac{1}{r^3} \bigg\rangle_{n\ell}
&=&
\frac{1}{2n}
\left(\frac{2}{na_0}\right)^3
\frac{2n}{2\ell (2\ell+1)(2\ell+2)} .
\label{eq:vev_pot_r3}
\end{eqnarray}
Similarly one may obtain matrix elements of potentials with $\log r$
by differentiating with respect to $u$.

\section{\boldmath Evaluation of $\Bigl< 
3 \frac{(\vec{r}\cdot \vec{S})^2}{r^2} - \vec{S}^2 
\Bigr>
$}
\label{App:DS}

We derive Eq.~(\ref{DefDS}) in this appendix.
According to the Wigner-Eckart theorem, a relation
\bea
\bra{\ell,\ell_z}\biggl(\frac{r_i r_j}{r^2}-\frac{1}{3}\delta_{ij}
\biggr)\ket{\ell,\ell_z'}
=f(\ell)\bra{\ell,\ell_z}\biggl[\frac{1}{2}\{L_i,L_j\}
-\frac{1}{3}\delta_{ij}\vec{L}^2
\biggr]\ket{\ell,\ell_z'}
\label{ApplyWEThm}
\eea
holds.
Setting $i=j=3$ and $\ell_z=\ell_z'=0$, we obtain
\bea
\bra{\ell,\ell_z=0}\biggl(\cos^2\theta-\frac{1}{3}
\biggr)\ket{\ell,\ell_z=0}
=f(\ell)\bra{\ell,\ell_z=0}\biggl(L_3^2
-\frac{1}{3}\vec{L}^2
\biggr)\ket{\ell,\ell_z=0}
.
\eea
Each side can be evaluated as
\bea
&&
({\rm l.h.s.})
=\frac{\int_{-1}^{+1}dx\, P_\ell(x)^2 x^2}
{\int_{-1}^{+1}dx\, P_\ell(x)^2}-\frac{1}{3}
=\frac{2\ell(\ell+1)}{3(2\ell-1)(2\ell+3)} ,
\\
&&
({\rm r.h.s.})
=-\frac{1}{3}\ell(\ell+1)\,f(\ell)
.
\eea
It follows that
\bea
f(\ell)=\frac{-2}{(2\ell-1)(2\ell+3)} .
\eea
Eq.~(\ref{ApplyWEThm}) implies that 
\bea
\bra{\ell,s;j,j_z}\biggl(\frac{r_i r_j}{r^2}-\frac{1}{3}\delta_{ij}
\biggr)\ket{\ell,s;j,j_z}
=f(\ell)\bra{\ell,s;j,j_z}\biggl[\frac{1}{2}\{L_i,L_j\}
-\frac{1}{3}\delta_{ij}\vec{L}^2
\biggr]\ket{\ell,s;j,j_z}
\nonumber\\
\eea
also holds.
Contracting both sides with $3S_iS_j$ yields
\bea
&&
\bra{\ell,s;j,j_z}\biggl[3 \frac{(\vec{r}\cdot \vec{S})^2}{r^2} - \vec{S}^2 
\biggr]\ket{\ell,s;j,j_z}
\nonumber\\&&
~~~~
=f(\ell)
\bra{\ell,s;j,j_z}\Bigl[
3(\vec{L}\cdot\vec{S})^2+\frac{3}{2}(\vec{L}\cdot\vec{S})
-\vec{S}^2\vec{L}^2
\Bigr]\ket{\ell,s;j,j_z}
\nonumber\\&&
~~~~
=f(\ell)
\Bigl[ 3 X_{LS}^2+\frac{3}{2}X_{LS}-\ell(\ell+1)s(s+1)\Bigr],
\eea
which coincides with Eq.~(\ref{DefDS}).

\section{\boldmath Derivation of $H^{us}$}
\label{appHUS}

In Eq.~(\ref{defX}),
we move all the momentum operators 
$\hat{\vec{p}}=-i\vec{\partial}_r$ to the right of 
the coordinate operators $\vec{r}$ and $r=|\vec{r}|$ using
commutation relations.
The commutation relations can be evaluated by taking
derivatives in coordinate space.
To keep track of the delta functions,  
we  regularize
the potentials as\footnote{
Even in dimensional regularization, such a regularization
is necessary, since delta functions in coordinate
space arise from
($d$-dimensional) Fourier transform of the relation
$q^2 \cdot {1}/{q^2} =1$.
}
\bea
V_S^{(d)}(r)\to -C_F\frac{\alpha_s}{r}
\left(\bar{\mu} r\right)^{2(\epsilon+u)} A(\epsilon)
\, ,~~~
V_O^{(d)}(r) \to \left(\frac{C_A}{2}-C_F\right)\frac{\alpha_s}{r}
\left(\bar{\mu} r\right)^{2(\epsilon+u)} A(\epsilon) \, .
\eea
We will send $u\to 0$ after Fourier transformation.
After a straightforward computation (which may be done easily using
an algebraic computational program), we obtain
\bea
&&
X=
-\frac{4 \alpha_s  A \, C_A \left(2 u^2+u (4 \epsilon
   -1)+\epsilon  (2 \epsilon -1)\right) (\bar{\mu}  r)^{2 (u+\epsilon )} 
}{m^2 r^3}\,r^ir^j\hat{p}^j\hat{p}^i
\nonumber\\&&
+\frac{\alpha_s ^3 A^3 C_A^3 (\bar{\mu}  r)^{6
   (u+\epsilon )}}{8 r}
+\frac{4 \alpha_s  A\,
   u (2 u+2 \epsilon -1) (C_A (u+1) (u+\epsilon )-C_F) (\bar{\mu} 
   r)^{2 (u+\epsilon )}}{m^2 r^3}
\nonumber\\&&
+\frac{\alpha_s ^2 A^2 C_A \left[C_A
   \{2 u^2\! + \!4 u (\epsilon \! + \!1)\! + \!2 \epsilon ^2\! + \!\epsilon \! + \!2 \}-4
   C_F (2 u^2\! + \!4 u \epsilon \! + \!u\! + \!2 \epsilon ^2\! + \!\epsilon
   \!-\! 1 )\right] (\bar{\mu}  r)^{4 (u+\epsilon )}}{2 m r^2}
\nonumber\\&&
+\frac{2 \alpha_s  A C_A 
   \left\{8 u^3+4 u^2 (4 \epsilon +1)+2 u \left(4 \epsilon ^2+6 \epsilon
   -3\right)+8 \epsilon ^2-6 \epsilon +1\right\} (\bar{\mu}  r)^{2
   (u+\epsilon )}}{m^2 r^3}\, i \vec{r}\cdot\hat{\vec{p}}
\nonumber\\&&
-\frac{2 \alpha_s  A C_A  
(2 u+2 \epsilon -1) (\bar{\mu}  r)^{2 (u+\epsilon )}}{m^2 r}\, \hat{\vec{p}}^{\,2}
.
\label{AppE:X}
\eea
In a similar manner, by evaluating $\vec{\partial}_r^{\,2} 
[{(\bar{\mu}  r)^{2   (u+\epsilon )}}/{r}]$,
one can show
\bea
&&
\frac{ (\bar{\mu}  r)^{2
   (u+\epsilon )}}{r^3}\, i \vec{r}\cdot\hat{\vec{p}} =
   \frac{iu}{r^3}(\bar{\mu}  r)^{2   (u+\epsilon )}
   \nonumber\\&&
   ~~~~~~~~~~~~~~~~~~~~
   +\frac{i}{4(u+\epsilon)-2}\Biggl(
   \biggl[ \hat{\vec{p}}^{\,2},
   \frac{(\bar{\mu}  r)^{2   (u+\epsilon )}}{r} \biggr]
   - \frac{(\bar{\mu}  r)^{2   (u+\epsilon )}}{r}
   \, \hat{\vec{p}}^{\,2}
   \Biggr).
\eea
We use this relation to eliminate $\vec{r}\cdot\hat{\vec{p}}$
from $X$.
The first term of Eq.~(\ref{AppE:X})
contributes only to the finite part of $H^{us}$, hence
we may use the relation
\bea
&&
\left[H_S^{(3)},\frac{C_F\alpha_s}{4mr}\Bigl(
-i\vec{r}\cdot\hat{\vec{p}}\Bigr)\right]
=-\frac{C_F\alpha_s}{4m^2}\left\{\frac{1}{r},\hat{\vec{p}}^{\,2}\right\}
+\frac{\pi C_F\alpha_s}{m^2}\delta^3(\vec{r})
\nonumber\\&&
~~~~~~~~~~~~~~~~~~~~~~~~~~~~~~~~~~~~~~~
+\frac{C_F\alpha_s}{2m^2r^3}\,r^ir^j\hat{p}^j\hat{p}^i+\frac{C_F^2\alpha_s^2}{4mr^2}
\, ,
\eea
which holds for $d=3$, to eliminate  $\frac{1}{r^3}\,r^ir^j\hat{p}^j\hat{p}^i$.
Note that the left-hand-side vanishes
inside the expectation value.

We take the Fourier transform of $X$ using the formulas 
in App.~\ref{app1}.
We insert the result in Eq.~(\ref{EUS2}) and
expand the local operators in $\epsilon$ and 
take the limit $u\to 0$.
Furthermore, we add
\bea
\left[H_S^{(d)},-\frac{2C_AC_F\alpha_s^2}{3m\pi r}\right]\Biggr|_{\rm F.T.}
=
-\frac{8 \alpha_s ^2 {C_A} {C_F} \left(p^2-p'^2\right)}{3 m^2
   q^2}
\eea
and obtain
eqs.~(\ref{EUS3})--(\ref{HUSren}).

\section{Angular Average}
\label{AppAngAve}

We explain how to take the sum over $\ell_z'$ in Eq.~(\ref{AngAve}).
We can do this in two different ways.

\subsection{Angular average using Clebsch-Gordan coefficients}
The spherical harmonic function for $\ell=1$ is given by
\begin{eqnarray}
Y_1^{\pm 1}(\theta,\phi)
&=&
\mp \sqrt{\frac{3}{8\pi}}\sin\theta e^{\pm i\phi}
=
\mp \sqrt{\frac{3}{8\pi}}\frac{x+iy}{r} ,
\\
Y_1^0(\theta,\phi) &=&\sqrt{\frac{3}{4\pi}} \cos\theta
=
\sqrt{\frac{3}{4\pi}}\frac{z}{r} .
\end{eqnarray}
Hence, one may write
\begin{eqnarray}
\hat{r}^i  \,  \hat{r}'^{\,i} 
&=&
\frac{1}{2} \left(\frac{x-iy}{r} \right) \left(\frac{x'+iy'}{r}\right)
+\frac{1}{2} \left(\frac{x+iy}{r}\right) \left(\frac{x'-iy'}{r}\right)
+ \left(\frac{z}{r}\right) \left(\frac{z'}{r'}\right)
\nonumber\\
&=&
\frac{4\pi}{3} \sum_{m=-1}^{+1} Y_1^{m\ast }(\hat{r}) \, Y_1^m(\hat{r}'),
\end{eqnarray}
where $\hat{r}$ denotes the unit vector in the direction of $\vec{r}$.

We can evaluate the following sum using 
the Clebsch-Gordan 
coefficients:
\begin{eqnarray}
\sum_{m'} T_{\ell m \ell' m'} 
&=&
\sum_{ m'} 
\langle \ell m|\, \hat{r}^i\, |\ell' m'\rangle
\langle \ell' m'|\,\hat{r}^i\, |\ell m\rangle
\nonumber\\
&=&
\frac{4\pi}{3}
\sum_{m'}
\sum_{i=\pm 1, 0}
\langle \ell m|\, Y_1^{i\ast } \, |\ell' m'\rangle
\langle \ell' m'|\,  Y_1^{i} \, |\ell m\rangle
\nonumber\\
&=&
\frac{4\pi}{3}
\sum_{m'}
\sum_{i=\pm 1, 0}
\bigg|
\bigg\langle 
\ell' m'
\bigg|
\begin{array}{cc}
1 & i\\
\ell & m
\end{array}
\bigg\rangle
\bigg|^2
\nonumber\\
&=&
\frac{4\pi}{3}
\bigg\{
\bigg | \bigg\langle 
\ell' ,0
\bigg|
\begin{array}{cc}
1 & 0\\
\ell & m 
\end{array}
\bigg\rangle \bigg|^2
+
\bigg | \bigg\langle 
\ell' ,1
\bigg|
\begin{array}{cc}
1 & +1\\
\ell & m 
\end{array}
\bigg\rangle \bigg|^2
+
\bigg | \bigg\langle 
\ell' ,-1
\bigg|
\begin{array}{cc}
1 & -1\\
\ell & m 
\end{array}
\bigg\rangle \bigg|^2
\bigg\}
\nonumber\\
&=&
\frac{\ell+1}{2\ell+1}\delta_{\ell', \ell+1} 
+\frac{\ell}{2\ell+1}\delta_{\ell',\ell-1}
\nonumber\\
&=&
\frac{\max(\ell,\ell')}{2\ell+1} 
\left(\delta_{\ell',\ell+1}+\delta_{\ell',\ell-1}\right)
\end{eqnarray}
where 
\bea
\bigg|
\begin{array}{cc}
1 & i\\
\ell & m 
\end{array}
\bigg\rangle 
\equiv
Y_{1}^i | \ell m\rangle
,
\eea
and we used
the selection rule for the $z$ component of the angular momentum:
\begin{eqnarray}
 \bigg\langle 
\ell' ,m'
\bigg|
\begin{array}{cc}
1 & i\\
\ell & m 
\end{array}
\bigg\rangle 
\propto \delta_{m', m+i} .
\end{eqnarray}
The Clebsch-Gordan coefficients are given as
\begin{eqnarray}
 \bigg\langle 
\ell+1 ,m
\bigg|
\begin{array}{cc}
1 & 0\\
\ell & m 
\end{array}
\bigg\rangle 
&=&
+\sqrt{\frac{3}{4\pi}} \,
\bigg[ \frac{(\ell-m+1)(\ell+m+1)}{(2\ell+1)(2\ell+3)}
\bigg]^{\frac{1}{2}}
, \\
 \bigg\langle 
\ell-1 ,m
\bigg|
\begin{array}{cc}
1 & 0\\
\ell & m 
\end{array}
\bigg\rangle 
&=&
+\sqrt{\frac{3}{4\pi}} \,
\bigg[ \frac{(\ell-m)(\ell+m)}{(2\ell-1)(2\ell+1)}
\bigg]^{\frac{1}{2}}
, \\
 \bigg\langle 
\ell+1 ,m+1
\bigg|
\begin{array}{cc}
1 & 1\\
\ell & m 
\end{array}
\bigg\rangle 
&=&
+\sqrt{\frac{3}{8\pi}} \,
\bigg[ \frac{(\ell+m+1)(\ell+m+2)}{(2\ell+1)(2\ell+3)}
\bigg]^{\frac{1}{2}}
, \\
 \bigg\langle 
\ell-1 ,m+1
\bigg|
\begin{array}{cc}
1 & 1\\
\ell & m 
\end{array}
\bigg\rangle 
&=&
-\sqrt{\frac{3}{8\pi}} \,
\bigg[ \frac{(\ell-m)(\ell-m-1)}{(2\ell-1)(2\ell+1)}
\bigg]^{\frac{1}{2}}
, \\
 \bigg\langle 
\ell+1 ,m-1
\bigg|
\begin{array}{cc}
1 & -1\\
\ell & m 
\end{array}
\bigg\rangle 
&=&
+\sqrt{\frac{3}{8\pi}} \,
\bigg[ \frac{(\ell-m+1)(\ell-m+2)}{(2\ell+1)(2\ell+3)}
\bigg]^{\frac{1}{2}}
, \\
 \bigg\langle 
\ell-1 ,m-1
\bigg|
\begin{array}{cc}
1 & -1\\
\ell & m 
\end{array}
\bigg\rangle 
&=&
-\sqrt{\frac{3}{8\pi}} \,
\bigg[ \frac{(\ell+m)(\ell+m-1)}{(2\ell-1)(2\ell+1)}
\bigg]^{\frac{1}{2}}
.
\end{eqnarray}

\subsection{Angular average using completeness relation of $Y_{\ell}^m$}
An alternative method is as follows.
One may easily see that
\begin{eqnarray}
T_{x,y}
&\equiv &
\sum_{m'} 
\langle x | \, \hat{r}^i\, |\ell' m'\rangle 
\langle \ell' m'| \, \hat{r}^i\,  | y \rangle
\nonumber\\
&=&
\hat{x}^i \hat{y}^i
\times
\sum_{m'=-\ell'}^{+\ell'} 
Y_{\ell'}^{m'}(\hat{x}) 
Y_{\ell'}^{m'*}(\hat{y})
\nonumber\\
&=&
\frac{2\ell'+1}{4\pi} 
\bigg\{ (\hat{x}\cdot\hat{y}) \,P_{\ell'}(\hat{x}\cdot \hat{y})\bigg\}
\nonumber\\
&=&
\frac{1}{4\pi} 
\bigg\{
(\ell'+1) P_{\ell'+1} (\hat{x} \cdot \hat{y})
+\ell' P_{\ell'-1}(\hat{x}\cdot\hat{y})
\bigg\}
\nonumber\\
&=&
\sum_{m'}
\bigg\{
\frac{\ell'+1}{2\ell'+3} 
\langle x |\ell'+1, m'\rangle 
\langle \ell'+1, m'|  y \rangle
+
\frac{\ell'}{2\ell'-1} 
\langle x |\ell'-1, m'\rangle 
\langle \ell'-1, m'|  y \rangle
\bigg\} ,
\nonumber\\
\end{eqnarray}
where we used the identities
\begin{eqnarray}
&&
\sum_{m=-\ell}^{+\ell} Y_\ell^m(\hat{x}) Y_\ell^{m*}(\hat{y}) =
\frac{2\ell+1}{4\pi} P_{\ell}(\hat{x}\cdot \hat{y}) ,
\\ &&
(2\ell+1) z P_\ell(z)= (\ell+1)P_{\ell+1}(z)+\ell P_{\ell-1}(z)
.
\end{eqnarray}
We may replace the external brackets $\bra{x}$,
$\ket{y}$ of $T_{x,y}$ by
$\bra{\ell, m}$, $\ket{\ell, m}$ and obtain
\begin{eqnarray}
\sum_{m'} 
\langle \ell , m| \hat{r}^i\, |\ell' m'\rangle 
\langle \ell' m'| \, \hat{r}^i |\ell, m\rangle
&=&
\frac{\ell'+1}{2\ell'+3} 
\delta_{\ell,\ell'+1}
+
\frac{\ell'}{2\ell'-1} 
\delta_{\ell,\ell'-1}
\nonumber\\
&=&
\frac{\ell}{2\ell+1} 
\delta_{\ell-1,\ell'}
+
\frac{\ell+1}{2\ell+1} 
\delta_{\ell+1,\ell'}
\nonumber\\
&=&
\frac{\max(\ell,\ell') }{2\ell+1} \bigg( \delta_{\ell-1,\ell'} +\delta_{\ell+1,\ell'}\bigg) .
\end{eqnarray}

\vspace*{10mm}

\begin{thebibliography}{0}


\bibitem{Pineda:1997bj} 
  A.~Pineda and J.~Soto,
  Nucl.\ Phys.\ Proc.\ Suppl.\  {\bf 64}, 428 (1998);
  N.~Brambilla, A.~Pineda, J.~Soto and A.~Vairo,
  Nucl.\ Phys.\ B {\bf 566}, 275 (2000).

\bibitem{Luke:1999kz} 
  M.~E.~Luke, A.~V.~Manohar and I.~Z.~Rothstein,
  Phys.\ Rev.\ D {\bf 61}, 074025 (2000).


\bibitem{Smirnov:2002pj}
  V.~A.~Smirnov,
  ``Applied asymptotic expansions in momenta and masses,''
  Springer Tracts Mod.\ Phys.\  {\bf 177} (2002) 1.

\bibitem{Smirnov:2004ym}
  V.~A.~Smirnov,
  ``Evaluating Feynman integrals,''
  Springer Tracts Mod.\ Phys.\  {\bf 211} (2004) 1.

\bibitem{Beneke:1997zp}
  M.~Beneke and V.~A.~Smirnov,
  Nucl.\ Phys.\ B {\bf 522} (1998) 321
  [hep-ph/9711391].

\bibitem{Brambilla:2004wf}
N.~Brambilla {\it et al.},  
arXiv:hep-ph/0412158;
  N.~Brambilla, {\it et al.},
  Eur.\ Phys.\ J.\ C {\bf 71} (2011) 1534.

\bibitem{Penin:2004ay}
  A.~A.~Penin, A.~Pineda, V.~A.~Smirnov and M.~Steinhauser,
  Nucl.\ Phys.\ B {\bf 699} (2004) 183
   [Erratum-ibid.\  {\bf 829} (2010) 398]
  [hep-ph/0406175].

\bibitem{Pineda:2006gx}
  A.~Pineda and A.~Signer,
  Phys.\ Rev.\ D {\bf 73} (2006) 111501
  [hep-ph/0601185].

\bibitem{Kiyo:2010jm}
  Y.~Kiyo, A.~Pineda and A.~Signer,
  Nucl.\ Phys.\ B {\bf 841} (2010) 231
  [arXiv:1006.2685 [hep-ph]].

\bibitem{Beneke:2014qea}
  M.~Beneke, Y.~Kiyo, P.~Marquard, A.~Penin, J.~Piclum, D.~Seidel and M.~Steinhauser,
  Phys.\ Rev.\ Lett.\  {\bf 112} (2014) 151801
  [arXiv:1401.3005 [hep-ph]].

\bibitem{Penin:2014zaa}
  A.~A.~Penin and N.~Zerf,
  JHEP {\bf 1404} (2014) 120
  [arXiv:1401.7035 [hep-ph]].
  
\bibitem{Ayala:2014yxa}
  C.~Ayala, G.~Cvetic and A.~Pineda,
  arXiv:1407.2128 [hep-ph].
  
\bibitem{Bazavov:2014soa}
  A.~Bazavov, N.~Brambilla, X.~G.~i.~Tormo, P.~Petreczky, J.~Soto and A.~Vairo,
  arXiv:1407.8437 [hep-ph].


\bibitem{Pineda:id}
A.~Pineda, Ph.D. Thesis;
A.~H.~Hoang, M.~C.~Smith, T.~Stelzer and S.~Willenbrock,
Phys.\ Rev.\ D {\bf 59}, 114014 (1999);
M.~Beneke,
Phys.\ Lett.\ B {\bf 434}, 115 (1998).


\bibitem{Pineda:1997hz}
  A.~Pineda and F.~J.~Yndurain,
  Phys.\ Rev.\ D {\bf 58} (1998) 094022;
  Phys.\ Rev.\ D {\bf 61} (2000) 077505;
  S.~Titard and F.~J.~Yndurain,
  Phys.\ Rev.\ D {\bf 49} (1994) 6007;
  Phys.\ Rev.\ D {\bf 51} (1995) 6348.

\bibitem{Brambilla:2001fw}
  N.~Brambilla, Y.~Sumino and A.~Vairo,
  Phys.\ Lett.\ B {\bf 513} (2001) 381;
  Phys.\ Rev.\ D {\bf 65} (2002) 034001.


\bibitem{Sumino:2001eh}
Y.~Sumino,
Phys.\ Rev.\ D {\bf 65}, 054003 (2002);
  S.~Necco and R.~Sommer,
  Nucl.\ Phys.\ B {\bf 622}, 328 (2002);
S.~Recksiegel and Y.~Sumino,
Phys.\ Rev.\ D {\bf 65}, 054018 (2002);

\bibitem{Pineda:2002se}
A.~Pineda,
J.\ Phys.\ G {\bf 29}, 371 (2003);
S.~Recksiegel and Y.~Sumino,
Eur.\ Phys.\ J.\ C {\bf 31}, 187 (2003);
 Y.~Sumino,
Phys.\ Rev.\  D {\bf 76}, 114009 (2007);
  N.~Brambilla, X.~Garcia i Tormo, J.~Soto and A.~Vairo,
  Phys.\ Rev.\ Lett.\  {\bf 105} (2010) 212001
   [Erratum-ibid.\  {\bf 108} (2012) 269903].

\bibitem{Anzai:2009tm} 
  C.~Anzai, Y.~Kiyo and Y.~Sumino,
Phys.\ Rev.\ Lett.\  {\bf 104}, 112003 (2010).

\bibitem{Kniehl:2002br}
  B.~A.~Kniehl, A.~A.~Penin, V.~A.~Smirnov and M.~Steinhauser,
  Nucl.\ Phys.\ B {\bf 635} (2002) 357.

\bibitem{Smirnov:2008pn}
  A.~V.~Smirnov, V.~A.~Smirnov and M.~Steinhauser,
  Phys.\ Lett.\  B {\bf 668}, 293 (2008).

\bibitem{Smirnov:2009fh} 
  A.~V.~Smirnov, V.~A.~Smirnov and M.~Steinhauser,
Phys.\ Rev.\ Lett.\  {\bf 104}, 112002 (2010).  

\bibitem{Kniehl:1999ud}
B.~A.~Kniehl and A.~A.~Penin,
Nucl.\ Phys.\ B {\bf 563}, 200 (1999).

\bibitem{Brambilla:1999xj}
  N.~Brambilla, A.~Pineda, J.~Soto and A.~Vairo,
  Phys.\ Lett.\ B {\bf 470} (1999) 215.

\bibitem{Beneke:2005hg}
  M.~Beneke, Y.~Kiyo and K.~Schuller,
  Nucl.\ Phys.\  B {\bf 714} (2005) 67;
  A.~A.~Penin, V.~A.~Smirnov and M.~Steinhauser,
  Nucl.\ Phys.\  B {\bf 716} (2005) 303.


\bibitem{Kiyo:2000fr} 
  Y.~Kiyo and Y.~Sumino,
  Phys.\ Lett.\ B {\bf 496}, 83 (2000);
  A.~H.~Hoang,
  hep-ph/0008102;
  A.~A.~Penin and M.~Steinhauser,
  Phys.\ Lett.\ B {\bf 538} (2002) 335.

\bibitem{Kiyo:2013aea}
  Y.~Kiyo and Y.~Sumino,
  Phys.\ Lett.\ B {\bf 730} (2014) 76
  [arXiv:1309.6571 [hep-ph]].

\bibitem{fn21}
A {Mathematica} package ${\it ``Wa"}$
for computing multiple sums using the
algorithms developed in \cite{Anzai:2012xw} is available
at 
\texttt{http://www.tuhep.phys.tohoku.ac.jp/$\sim$program/} 
with examples and instructions.

\bibitem{Anzai:2012xw}
  C.~Anzai and Y.~Sumino,
  J.\ Math.\ Phys.\ {\bf 54} (2013) 033514
  [arXiv:1211.5204 [hep-th]].


\bibitem{Beneke:2013jia}
  M.~Beneke, Y.~Kiyo and K.~Schuller,
  arXiv:1312.4791 [hep-ph].
  

\bibitem{Beneke:2007pj}
  M.~Beneke, Y.~Kiyo and A.~A.~Penin,
  Phys.\ Lett.\ B {\bf 653} (2007) 53
  [arXiv:0706.2733 [hep-ph]];
  M.~Beneke, Y.~Kiyo and K.~Schuller,
  Phys.\ Lett.\ B {\bf 658} (2008) 222
  [arXiv:0705.4518 [hep-ph]].


\end{thebibliography}

\end{document}